\documentclass[twocolumn, switch]{article} 

\usepackage{preprint}

\usepackage{amsmath, amsthm, amssymb, amsfonts}

\usepackage[numbers,square]{natbib}
\usepackage{natbib}

\usepackage[utf8]{inputenc}	
\usepackage[T1]{fontenc}	
\usepackage{xcolor}		
\usepackage[colorlinks = true,
            linkcolor = purple,
            urlcolor  = blue,
            citecolor = cyan,
            anchorcolor = black]{hyperref}	
\usepackage{booktabs} 		
\usepackage{nicefrac}		
\usepackage{microtype}		
\usepackage{lineno}		
\usepackage{float}			

\usepackage{lipsum}		

\usepackage{newfloat}
\DeclareFloatingEnvironment[name={Supplementary Figure}]{suppfigure}
\usepackage{sidecap}
\sidecaptionvpos{figure}{c}

\usepackage{titlesec}
\titlespacing\section{0pt}{12pt plus 3pt minus 3pt}{1pt plus 1pt minus 1pt}
\titlespacing\subsection{0pt}{10pt plus 3pt minus 3pt}{1pt plus 1pt minus 1pt}
\titlespacing\subsubsection{0pt}{8pt plus 3pt minus 3pt}{1pt plus 1pt minus 1pt}

\usepackage{tikz,xcolor,hyperref}

\definecolor{lime}{HTML}{A6CE39}
\DeclareRobustCommand{\orcidicon}{
	\begin{tikzpicture}
	\draw[lime, fill=lime] (0,0)
	circle [radius=0.16]
	node[white] {{\fontfamily{qag}\selectfont \tiny ID}};
	\draw[white, fill=white] (-0.0625,0.095)
	circle [radius=0.007];
	\end{tikzpicture}
	\hspace{-2mm}
}
\foreach \x in {A, ..., Z}{\expandafter\xdef\csname orcid\x\endcsname{\noexpand\href{https://orcid.org/\csname orcidauthor\x\endcsname}
			{\noexpand\orcidicon}}
}

\usepackage{derivative}
\usepackage{siunitx} 
\usepackage{subcaption} 
\usepackage{algorithm}
\usepackage{algpseudocodex}
\usepackage[export]{adjustbox}
\usepackage{listofitems}

\algnewcommand\CallTempl[3]{%
  \textproc{#1}%
  \ifstrempty{#2}{}{\textlangle{#2}\textrangle}%
  \ifstrempty{#3}{}{(#3)}
}

\newcommand{\imagetable}[1]{%
  \setsepchar{,}  
  \readlist{\images}{#1}

  \begin{tabular}{llll}
    & \mcc{Density} & \mcc{Ionization fraction} & \mcc{Temperature} \\
    \rotLabel{$1024^3$} &
      \includegraphics[width=.3\linewidth,valign=m]{\images[1]} &
      \includegraphics[width=.3\linewidth,valign=m]{\images[2]} &
      \includegraphics[width=.3\linewidth,valign=m]{\images[3]} \\
    \rotLabel{$2048^3$} &
      \includegraphics[width=.3\linewidth,valign=m]{\images[4]} &
      \includegraphics[width=.3\linewidth,valign=m]{\images[5]} &
      \includegraphics[width=.3\linewidth,valign=m]{\images[6]} \\
    \rotLabel{$4096^3$} &
      \includegraphics[width=.3\linewidth,valign=m]{\images[7]} &
      \includegraphics[width=.3\linewidth,valign=m]{\images[8]} &
      \includegraphics[width=.3\linewidth,valign=m]{\images[9]} \\
  \end{tabular}
}

\newcommand{\imagetablefour}[1]{%
  \setsepchar{,}
  \readlist{\images}{#1}

  \begin{tabular}{lll}
    & \mcc{No Source Merging} & \mcc{Source Merging} \\
    \rotLabel{$z=6.9$} &
      \includegraphics[width=.3\linewidth,valign=m]{\images[1]} &
      \includegraphics[width=.3\linewidth,valign=m]{\images[3]} \\
    \rotLabel{$z=5.8$} &
      \includegraphics[width=.3\linewidth,valign=m]{\images[2]} &
      \includegraphics[width=.3\linewidth,valign=m]{\images[4]} \\
  \end{tabular}
}

\newcommand{\cTwoRay}{C$^2$-Ray}
\newcommand{\pycTwoRay}{pyC$^2$Ray}
\newcommand{\Arc}{\textsc{Arc}}
\newcommand{\safetyF}{\ensuremath{N_{\textrm{min}}^{\textrm{ray}}}}
\newcommand{\fctp}{\textsc{filterCopyTransformParticles}\textlangle{$k$}\textrangle}
\newcommand{\rotLabel}[1]{\rotatebox[origin=c]{90}{#1}}
\newcommand{\mcc}[1]{\multicolumn{1}{c}{#1}}
\newcommand{\luminosity}{\dot{N}_\gamma}
\newcommand{\Vbox}{V_{\mathrm{box}}}

\DeclareSIUnit \h {\ensuremath{\mathit{h}}}
\DeclareSIUnit \parsec {pc}
\DeclareSIUnit \solarmass {\textup{M}_\odot}
\title{Nyx-RT: Adaptive Ray Tracing in the Nyx Hydrodynamical Code}

\usepackage{xwatermark}
\newwatermark[firstpage,color=gray!60,angle=90,scale=0.32, xpos=3.9in,ypos=0]{\href{https://doi.org/}{\color{gray}{Preprint doi}}}
\newwatermark[firstpage,color=gray!90,angle=0,scale=0.28, xpos=0in,ypos=-5in]{*correspondence: \texttt{nmarshak@sci.utah.edu}}

\usepackage{authblk}

\author[1,3\thanks{\texttt{nmarshak@sci.utah.edu}}]{Nathan X. Marshak\orcidA{}}
\author[2,3]{Kathlynn Simotas\orcidB{}}
\author[3]{Zarija Luki\'c\orcidC{}}
\author[3,4]{Hyunbae Park\orcidD{}}
\author[5]{James Ahrens\orcidE{}}
\author[1]{Chris.~R.~Johnson\orcidF{}}

\affil[1]{Scientific Computing and Imaging Institute, University of Utah, Salt Lake City, UT 84112, USA}
\affil[2]{Department of Physics, University of California, Santa Barbara, CA 93106, USA}
\affil[3]{Computational Cosmology Center, Lawrence Berkeley National Laboratory, Berkeley, CA 94720, USA}
\affil[4]{Center for Computational Sciences, University of Tsukuba, 1-1-1 Tennodai, Tsukuba, Ibaraki 305-8577, Japan}
\affil[5]{Information Science \& Technology Institute, Los Alamos National Laboratory, Los Alamos, NM 87545, USA}

\begin{document}

\twocolumn[ 
  \begin{@twocolumnfalse} 

\maketitle

\begin{abstract}
	Numerical methods for radiative transfer play a key role in modern-day astrophysics and cosmology, including study of the inhomogeneous reionization process.
	In this context, ray tracing methods are well-regarded for accuracy but notorious for high computational cost.
	In this work, we extend the capabilities of the \textsc{Nyx} $N$-body / hydrodynamics code, coupling radiation to gravitational and gas dynamics.
  We formulate adaptive ray tracing as a novel series of filters and transformations that can be used with \textsc{AMReX} particle abstractions, simplifying implementation and enabling portability across Exascale GPU architectures.
	To address computational cost, we present a new algorithm for merging sources, which significantly accelerates computation once reionization is well underway.
	Furthermore, we develop a novel prescription for geometric overlap correction with low-density neighbor cells.
	We perform verification and validation against standard analytic and numerical test problems.
	Finally, we demonstrate scaling to up to 1024 nodes and 4096 GPUs running multiphysics cosmological simulations, with 4096$^3$ Eulerian gas cells, 4096$^3$ dark matter particles, and ray tracing on a $1024^3$ coarse grid.
	For these full cosmological simulations, we demonstrate convergence in terms of reionization history and post-ionization Lyman-$\alpha$ forest flux.
\end{abstract}
\vspace{0.35cm}

  \end{@twocolumnfalse} 
] 



\section{Introduction}
Reionization, the transformation of the intergalactic medium (IGM) from a cold, neutral gas to an ionized hot plasma, is one of the most characteristics phase transitions in cosmic history, and it remains a topic of central importance to cosmology \citep{aghanim2020planck,gnedin2022modeling, 10.1093/mnras/stw3026}.
Also known as ``cosmic noon", reionization starts with the dawn of the luminous universe with the birth of early stars, galaxies, and accreting black holes.
These luminous objects and their interaction with the IGM dramatically altered the state of baryons and shaped the subsequent evolution of large-scale structure.
However, despite decades of observational and theoretical progress, key questions about the process of reionization remain unresolved \citep{gnedin2022modeling,Bouwens_2016,cain_short_2021,wolfson2025forecasting}. Some of the most pressing include:
\begin{itemize}
	\item When did the hydrogen and helium reionization phase transitions occur?
	\item How clumpy and inhomogeneous was reionization?
	\item Which sources are most responsible for reionization?
\end{itemize}
These and other persistent questions cannot be answered with only direct observation or analytical modeling beyond linear scales.
Reionization is intrinsically non-linear, non-local, and highly inhomogeneous \citep{Becker_2015}.
Thus, numerical simulations, particularly large-scale hydrodynamical simulations, which evolve the distribution of matter in the IGM, have become indispensable research tools to connect theoretical models with observables.
\par
With the influx of high-quality data from new and upcoming observations made with (among others) JWST, LOFAR, HERA, SKA, along with improved statistical measurements (e.g.
1D and 3D flux power spectra, higher-order correlation functions, cross-correlations), the potential for constraining properties of reionization has grown \citep{robertson2022galaxy}.
Extracting meaningful results from this data requires simulations that faithfully represent the physics of the IGM, up to and including the impact of radiation and inhomogeneous reionization as determined by ionizing sources~\citep{kannan2022introducing}.
\par
Despite years of work, large-scale cosmological hydrodynamics simulations like \textsc{Nyx} still use semi-numerical methods or moment methods to model reionization.
While computationally efficient, these approaches trade off physical fidelity and cannot entirely capture the complex, non-linear processes of reionization. This limitation becomes increasingly problematic as observational precision grows.
\par
\textsc{Nyx-RT} addresses this by coupling physically accurate ray tracing to hydrodynamics in \textsc{Nyx}, overcoming the fundamental limitations of semi-numerical models used in past \textsc{Nyx} studies.
Our approach balances computational cost with physical fidelity, enabling a significant improvement over existing \textsc{Nyx} functionality while maintaining scalability, as demonstrated by successful runs on 4096 GPUs at $4096^3$ resolution.
\par
Previously, various radiative transfer methods been used to directly relate models of reionization to observables in simulations.
These methods have a wide range and degree of simplifying assumptions made due to the computational difficulty of solving the six-dimensional radiative transport equation \citep{wu_accuracy_2021}.
Semi-numerical approaches, including methods based on the excursion-set formalism, are generally considered to be the most computationally efficient, and some models can execute in minutes on a single CPU.~\citep{zahn_simulations_2007,zahn_comparison_2011,mesinger_21cmfast_2011}
However, results can be limited in physical fidelity due to the nature of semi-numerical methods: they can be thought of as a compromise between simple analytic solutions and large-scale ``first-principles'' numerical simulation.
Thus, they do not necessarily include the physics needed to exactly capture the position and structure of ionization fronts, and some assume uniform reionization.
As a result, simulation results are used to validate~\citep{almgren_nyx_2013} and/or calibrate~\citep{battaglia_reionization_2013} semi-numerical models.~\citep{zahn_comparison_2011}
\par
Moment-based methods are numerical simulation techniques derived from taking the moments of the radiative transfer equation~\citep{norman_simulating_1998, gnedin_multi-dimensional_2001, aubert_radiative_2008,finlator_new_2009,kannan_arepo-rt_2019}.
These approaches have the advantage of scaling that is independent of the number of sources: $\mathcal{O}(N\log N)$  where $N$ is the number of cells~\citep{gnedin_multi-dimensional_2001}.
However, most methods in this class approximate higher angular moments using a closure relation such as OVTET~\cite{gnedin_multi-dimensional_2001} or M1~\citep{levermore_relating_1984, aubert_radiative_2008},
leading to a ``diffusive'' approximation~\citep{hartley_arc_2019, mellema_c2-ray_2006} that often has difficulty resolving sharp features such as shadows cast by opaque absorbers, ``dense, self-shielding clumps''~\citep{wu_accuracy_2021}, or ``enhanced molecule formation due to shocks''~\citep{iliev_cosmological_2009}.
In addition, moment-based methods usually assume a reduced speed of light, which introduces a nonphysical dependence between the adopted speed of light and the neutral hydrogen fraction \citep{wu_accuracy_2021}.
\par
Ray tracing is generally considered the most accurate and best able to capture the inhomogeneous evolution of reionization in comparison to semi-numerical and moment-based methods, provided that the pixelization (i.e. discretization) in angular coordinates is of sufficient resolution~\citep{wu_accuracy_2021}.
However, it generally comes at a greater computational cost.
In addition to the large number of rays and ray marching steps required, scaling is linear in the number of sources unless special considerations are made, such as source grouping with a tree-based data structure~\citep{Razoumov2002,Hasegawa2010,Okamoto2012,wunsch_tree-based_2018, grond_trevr_2019}.
\par
There are some techniques that do not fit neatly into the three categories of semi-numerical, moment-based, and ray tracing.
For instance,~\cite{finlator_new_2009} developed a moment-based method that uses long characteristics ray tracing to compute the Eddington tensor.~\cite{petkova_novel_2011} and the \textsc{Traphic} family of codes~\citep{pawlik_traphic_2008,pawlik_multifrequency_2011,raicevic_effect_2014} use a cone-based geometry for photon transport.
\cite{molaro_artist_2019} proposes a novel numerical method based on spherical shells.
In addition, there are several works that use the Monte Carlo method~\citep{ciardi_cosmological_2001,cantalupo_radamesh_2011}, including the \textsc{Crash} code~\citep{maselli_crash_2003,maselli_crash2_2009,partl_enabling_2011,holst_crash_2011}.

\par
Implementing radiative transfer at the largest scales (even approaching an order of magnitude of the size of modern surveys) remains challenging \citep{wu_accuracy_2021, gnedin2014cosmic,ocvirk_cosmic_2020,Anninos_1997}. Incorporating ray tracing based radiative transfer (RT) into cosmological-scale hydrodynamical simulations is extremely expensive, severely limiting the ability to perform parameter-space explorations or systematic studies of reionization histories, source models, and radiative feedback \citep{kannan2022introducing,gnedin2022modeling, 10.1093/mnras/stw3026}.
New approaches coupled with more modern computing infrastructures that encourage the use of high performance and parallel computing, emulation, and other techniques are needed to push coupled radiative transfer in hydrodynamics simulations into the regime of usefulness to large scale simulations and studies requiring them.
\par
To make better use of modern high performance computing paradigms, several previous authors have explored GPU-based methods for radiative transfer and hydrodynamics in the context of reionization.
Most~\citep{aubert_reionization_2010,rosdahl_ramses-rt_2013,bauer_hydrogen_2015,aubert_emma_2015,Ocvirk_2016, ocvirk_cosmic_2020} couple hydrodynamics to a moment-based method for radiation transport.
\cite{bauer_hydrogen_2015}, in addition to a moment-based solver, also implements the cone-based advection method of~\cite{petkova_novel_2011} for radiation.
Several prior works explore GPU ray tracing for reionization problems, including~\cite{thomson_ray-traced_2018}, who develop a GPU ray tracing technique that acts as a post-process for particle-based SPH simulation data.
In addition, the \Arc{} code~\citep{hartley_arc_2019} implements GPU ray tracing for uniform grid-based data using CUDA.
Furthermore, \pycTwoRay{} is a recent GPGPU implementation of \cTwoRay{}~that advances rays using \textsc{ASORA}, a novel short characteristics method, implemented in CUDA, that traverses the volume using octahedral wavefronts.~\citep{Hirling_2024}
\par
In response to the abilities of modern supercomputers and growing need to study reionization on both the largest and smallest scales, the Nyx code has emerged as a particularly suitable code for cosmological simulations aimed at modeling reionization and the Ly-$\alpha$ forest.
Nyx is a massively parallel, Eulerian, grid-based adaptive mesh code built on the \textsc{AMReX} framework \citep{almgren_nyx_2013,Lukic_2014}.
It has been used extensively to simulate the IGM and generate synthetic Ly-$\alpha$ forest spectra, emulate the epoch of reionization, and understand the impact of various reionization histories on observables \citep{Lukic_2014,onorbe_inhomogeneous_2019, chabanier2023modelling, gonzalez2025using,wolfson2025forecasting, Walther_2025}.
\par
Nyx is optimized for modern, high performance computing environments, being built on top of the AmReX framework \citep{almgren_nyx_2013, AMReX_JOSS}.
AmReX has Exascale class parallelism and maximum portability across architectures and GPUs. Its design emphasizes scalability, resolution, and compatibility with modern high-performance computing architectures such as the computing facilities at NERSC, OLCF, and ALCF, which positions it as an ideal platform on which to implement a fully coupled radiation-hydrodynamics (rad-hydro) scheme capable of handling cosmological volumes \cite{AMReX_JOSS}.
Prior to this work, the \textsc{21cmfast} code~\citep{mesinger_21cmfast_2011} or similar semi-numerical models were used with Nyx to simulate the impact of inhomogeneous reionization~\citep{onorbe_inhomogeneous_2019,park2024b,Doughty_2025}.
\par
We now present a ray tracing system, integrated with Nyx, that uses adaptive ray tracing to simulate reionization events.
Our work includes a novel source merging scheme that accelerates computation, and an extension of the geometric overlap correction in past work~\cite{wise_enzomoray_2011, hartley_arc_2019} that overcomes numerical difficulties caused by low-density neighbor cells.
\par
For adaptive ray tracing specifically, distributed GPU implementation at Exascale remains a challenge.
Unlike many scientific GPU codes, our implementation needs to be vendor-agnostic.
Performance portability is a must for deployability across diverse range of vendors and architectures, including AMD, Intel, and NVIDIA.
Our solution is a novel formulation of adaptive ray tracing, expressed as a series of \textsc{AMReX} particle filters and transformations.
This also simplifies implementation and enables code re-use:
We can express a complex, distributed algorithm using concise, self-contained functions that are given as input to \textsc{AMReX} primitives which abstract away much of the complexity of distributed GPGPU programming, in addition to platform-specific details.

\par


\par
Given the growing volume and precision of high-redshift datasets (e.g. from JWST, DESI, future 21-cm experiments), there is a pressing need for methods that combine physical fidelity (e.g.
ray tracing based radiative transfer) with computational efficiency \citep{gnedin2022modeling, wu_accuracy_2021, kannan2022introducing, robertson2022galaxy}.
Doing so would enable systematic investigations of how different reionization histories and source populations affect observables (e.g. Ly-$\alpha$ forest power spectra, 21-cm statistics, quasar absorption spectra), and thus strengthen the connection between theory and data.
Incorporating radiative transfer via ray tracing in a GPU-enabled, Exascale-ready code like Nyx provides a promising avenue for this ongoing work.
\par
Our paper is organized as follows.
Section~\ref{sec:method} describes how we implemented our GPU ray tracer and how we coupled it to Nyx's $N$-body/hydro solvers.
In Section~\ref{sec:eval}, we present our computational performance measurements.
We discuss these results in Section~\ref{sec:large_scale_simulation} and conclude in Section~\ref{sec:conc}.

\section{Method}\label{sec:method}

\subsection{Review of Ray Tracing Methods}
\subsubsection{Short and long characteristics}\label{sec:past_short_long_char}
Ray tracing techniques are often classified as \emph{short characteristics} or \emph{long characteristics} methods.
Long characteristics methods cast an independent ray to each cell, which is accurate but require a large number of repeated cell traversals.
Short characteristics techniques sacrifice some accuracy for speed by traversing along line segments and making an approximation to compute the column density at the beginning of the segment.
Some short characteristics codes for grid-based data follow a pattern similar to~\cite{raga_3d_1999}: Starting from the source cell, column densities are computed with multiple axis-aligned sweeps of the volume.~\citep{lim_3d_2003, mellema_c2-ray_2006}
This work focuses on a different notion of short characteristics: \emph{adaptive ray tracing}, first introduced by~\cite{abel_adaptive_2002}.
Adaptive ray tracers~\citep{wise_enzomoray_2011,hartley_arc_2019,cain_short_2021} successively split rays cast from the source into child sub-rays, using the ``HEALPix'' multi-resolution discretization of a sphere~\citep{gorski_healpix_2005} to determine the origin of each sub-ray.

\subsubsection{Geometric overlap correction}\label{sec:geom_correction_literature}
For adaptive ray tracing, \cite{wise_enzomoray_2011} reduce spatial discretization artifacts by considering the volume swept by each ray and modulating per-cell incident flux by the amount of ray-cell overlap.
\cite{hartley_arc_2019} extend this scheme and enforce photon conservation by depositing flux that falls outside the cell into its nearest neighbor.

\subsubsection{Post-processing codes and coupled codes}\label{sec:post_and_coupled}
Works such as~\cite{mellema_c2-ray_2006} and~\cite{hartley_arc_2019} are examples of ray tracers that are presented as a ``standalone'' post-processors that apply radiation effects to the output of an $N$-body and hydrodynamics simulator.
While post-processing is computationally efficient, radiation and gas dynamics cannot co-evolve, therefore results may ``fail to self-consistently account for processes leading to the formation of ionizing sources, and the co-evolution of the source and sink populations.''~\citep{molaro_artist_2019}
\par
An alternative, with higher physical fidelity, is to couple ray tracing to hydrodynamics routines using techniques such as operator splitting, for instance the short-characteristics $C^2$-ray algorithm coupled to the \textsc{Capreole} ~\citep{mellema_dynamical_2006} hydrodynamics code, or the \textsc{Moray} HEALPix-based ray tracer coupled to the adaptive mesh refinement \textsc{Enzo} code.~\citep{wise_enzomoray_2011}.
The \textsc{iVine} code uses plane-parallel ray tracing in conjunction with SPH for fluid simulation~\citep{gritschneder_ivine_2009}.
The Cosmological Radiative Transfer Comparison Project~\citep{iliev_cosmological_2009} enumerates seven more.
Finally, there are numerous examples of coupled codes that use a radiative transfer solution other than ray tracing.~\citep{petkova_implementation_2009,pawlik_first_2013,pawlik_aurora_2017,park2016,pawlik_spatially_2015,rosdahl_ramses-rt_2013,rahmati_evolution_2013,norman_fully_2015}

\begin{figure*}
	\begin{center}
		\includegraphics[width=0.8\textwidth]{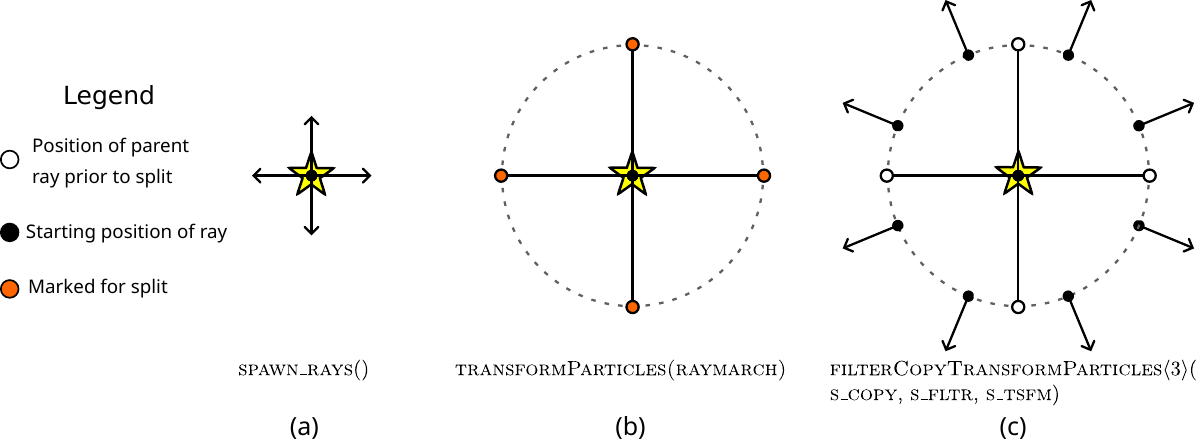}
	\end{center}
	\caption{Schematic diagram of our adaptive ray tracing implementation (Sec.~\ref{sec:flux_comp}, Alg.~\ref{alg:trace_rays}), ignoring domain boundaries, box boundaries and early ray termination. (a) At the source position, initialize 12 particles, each corresponding to a pixel on the HEALPix sphere at the 0th refinement level. (b) March and deposit flux until at least one stopping condition is satisfied. In this example, marching stops when rays reach the splitting radius $r_{\textrm{split}}$. (c) Filter particles in order to find the subset that is marked for splitting. Split each parent ray into four children, each representing a pixel at the next HEALPix refinement level on a sphere of radius $r_{\textrm{split}}$. }\label{fig:raysplit}
\end{figure*}

\begin{figure*}
	\begin{center}
		\includegraphics[width=0.95\textwidth]{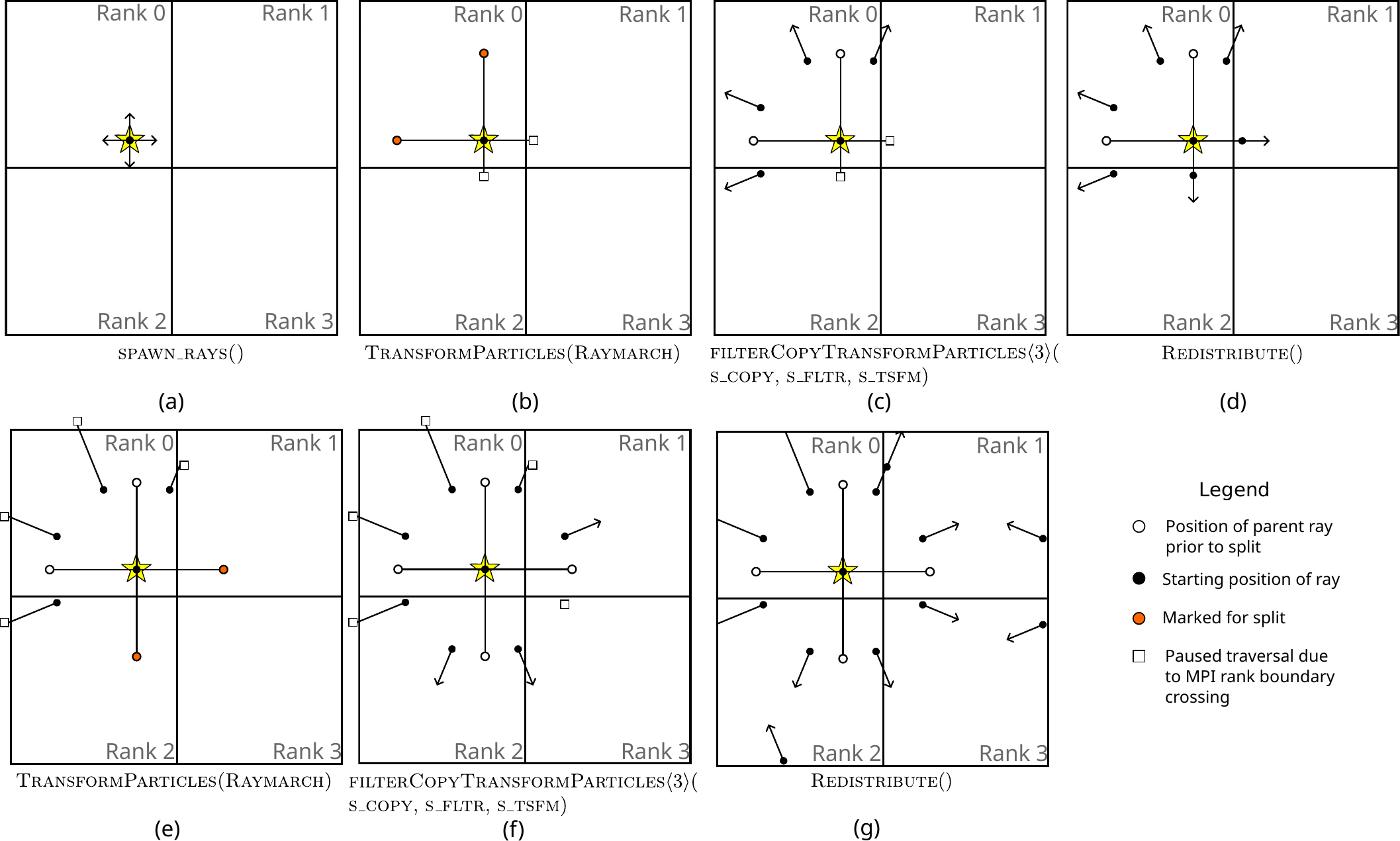}
	\end{center}
	\caption{Example ray traversal (Sec.~\ref{sec:flux_comp}, Alg.~\ref{alg:trace_rays}) that illustrates handling of subvolume boundaries and periodic BCs. (a) Rays are initialized like in Fig~\ref{fig:raysplit}. (b) Raymarch until either a boundary is reached or the splitting criterion is satisfied. In the latter case, mark the ray for splitting. (c) Split the particles that are marked for splitting. (d) Call \textsc{Redistribute()} in order to send rays that have crossed a subvolume boundary to the correct MPI rank. (e) Raymarch again, like in (b). (f) Split rays again. In this example, one of the newly-created rays starts inside the subvolume owned by Rank 3, even though its parent was on Rank 1. (g) Call \textsc{Redistribute()} again. This time, in addition to handling rays that have crossed a subvolume boundary, rays that have crossed a \emph{domain} boundary wrap around as a result of periodic BCs.}\label{fig:boundary}
\end{figure*}

\subsection{Flux computation}\label{sec:flux_comp}
\par
At each cell traversed, a photon-conserving formula~\citep{Abel199966, mellema_c2-ray_2006} is used to compute $\Gamma$, which is the ionizing flux per H atom:
\begin{equation}\label{eq:lut_eqn}
	 \Gamma = \int_{\nu_0}^{\infty} \frac{L_\nu e^{-\tau_\nu}}{h\nu} \frac{1 - e^{-\Delta \tau_\nu}}{n_{HI} V_{shell}} \odif{\nu}
\end{equation}
where $v_0$ is the H ionization threshold frequency, $L_\nu$ is the luminosity at frequency $\nu$, $\tau_\nu$ the frequency-dependent optical depth up to the cell, $\Delta \tau_\nu$ the cell's optical depth, and $n_{HI}$ the neutral hydrogen number density.
Note that $V_{shell} = \frac{4\pi}{3}[(R + h)^3 - R^3]$ for a cell of width $h$ at radius $R$ from the source.
\par
Let $x$ denote the ionization fraction, and $\sigma_{\nu}$ the photoionization cross section at frequency $\nu$. 
Then the optical depth $\tau$ along a ray is given by:
\begin{equation}
	\label{eq:optical_depth} \tau_{\nu}(R) = \int_0^R \sigma_{\nu} x(r)n_{HI}(r) \odif{r}
\end{equation}
Following the approach of~\cite{mellema_c2-ray_2006}, the solution to Eq.~\ref{eq:lut_eqn} can be precomputed as a function of $\tau_0$ (optical depth at the ionization threshold frequency).
A lookup table allows us to sidestep the need for per-ray frequency bins, or to have separate rays carry photons at different frequencies.
\subsection{Time integration}\label{sec:inner_it}
Once $\Gamma$ from each point source is accumulated, we must solve the ODE in Eq.~\ref{eq:chem_ode} at each cell, where $x$ is the ionized fraction, $n_e = xn_{HI}$ is the local electron density, $C_H$ is the collisional ionization coefficient for hydrogen, and $\alpha_H$ is the recombination coefficient.
\begin{equation}
	\label{eq:chem_ode}
	\odv{x}{t} = (1-x)(\Gamma + n_eC_H) - xn_e\alpha_H
\end{equation}
For time integration we use the semi-analytic method of~\cite{mellema_c2-ray_2006}, which introduces \emph{time averaging}, enabling larger timesteps while retaining accuracy and numerical stability.

\subsection{Iterative nonlinear solve}\label{sec:nonlinear_solve}
Continuing in the vein of~\cite{mellema_c2-ray_2006}, a numerical solution to the system in Eqs.~\ref{eq:lut_eqn}, \ref{eq:chem_ode}, and \ref{eq:optical_depth} can be found by fixing $x$ and evaluating Eqs.~\ref{eq:optical_depth} and \ref{eq:lut_eqn} to compute $\Gamma$, then fixing $\Gamma$ and updating the value of $x$ via time integration of Eq.~\ref{eq:chem_ode}, and iterating until convergence.

\subsection{Adaptive ray tracing}\label{sec:tracing}
In order to compute $\Gamma$, we must compute the optical depth from each source to each cell.
Per the adaptive ray tracing scheme (Sec.~\ref{sec:past_short_long_char}), traversal starts from each source with twelve raycasts, with each ray corresponding to a pixel at the coarsest level of the \textsc{HEALPix} discretization.
Each ray marches forward until the splitting criterion is satisfied:
\begin{equation}\label{eq:split_criterion}
	\Omega_c / \Omega(\ell) < \safetyF
\end{equation}
Assuming each cell is a perfect cube with side length $\Delta x$, $\Omega_c \approx (\Delta x)^2 / R^2$ is the solid angle subtended by a cell from the perspective of a point source at distance $R$.
$\Omega(\ell) = 4\pi / (12 \times 4^\ell)$ is the solid angle subtended by a single HEALPix pixel at refinement level $\ell$. $\safetyF \geq 1$ is the prescribed minimum number of rays per cell, also known as the \emph{safety factor} or \emph{splitting rate}.
If Eq.~\ref{eq:split_criterion} is satisfied, each ray splits into four HEALPix subpixels, following the approach of~\cite{abel_adaptive_2002}.
%
\par
The process repeats until the all rays terminate.
Rays may be terminated because they exceed a maximum optical depth threshold or a maximum number of splits.
\subsubsection{Periodic BCs}\label{sec:adaptive_periodic}
In the case of periodic boundary conditions, rays do not terminate once they cross a domain boundary.
Instead they wrap around, terminating when their travel distance exceeds $\sqrt{3}L_{box}$, where $L_{box}$ is the length of one side of the domain box, assuming it is a perfect cube.

\subsubsection{\textsc{AMReX} particle filtering and transformation}\label{sec:amrex_adaptive_rt}
We formulate adaptive ray tracing as a series of filter and transform operations with the help of the following \textsc{AMReX} primitives:
\begin{enumerate}
	\item \textsc{transformParticles}\label{tsfm}
	\item \textsc{filterParticles}\label{fltr}
	\item  \textsc{filterCopyTransformParticles}{\textlangle{$k$}\textrangle} \label{fltrCpyTsfm}
\end{enumerate}

\textsc{transformPraticles} takes a function $f$ as input, and applies $f$ to each particle.
\textsc{filterParticles} takes a predicate $q$ as input, and filters the set of particles $P$ based on whether $q(p)$ evaluates to \texttt{True} or \texttt{False} for each $p \in P$.
Both are provided by the base \textsc{AMReX} library.
\par
\fctp{} is an abstraction that comes from the \textsc{WarpX} Particle-In-Cell code~\citep{Fedeli_2022, Vay2025}, originally used to model a hydrogen atom ionizing and ``splitting'' into a proton and electron. Its inputs are an integer template parameter $k$, a copy function $c$, a filtering predicate $r$, and a transformation function $g$.
For each $p \in P$, if $r(p)$ evaluates to true, $k$ new particles are created, using $c$ to initialize the data in the newly-created particles at copy time.
Finally, $g$ is applied to the original particle.
\par
Sec.~\ref{sec:user_defined} details our input functions and predicates. Sec.~\ref{sec:rt_loop} and Figs.~\ref{fig:raysplit}, \ref{fig:boundary} explain how we use \ref{tsfm}, \ref{fltr}, and \ref{fltrCpyTsfm} in our ray tracing loop.

\subsubsection{User-defined filter and transform functions}\label{sec:user_defined}
\begin{algorithm}
\begin{algorithmic}
	\Function{raymarch}{$p$}
	\While{$(\neg \textsc{need\_split}(p) ) \land (p \in D_L) \land (p.s < \sqrt{3}L_{\textrm{box}})$}
	\State \Call{advance\_one\_cell}{$p$}
	\State Compute flux (Sec.~\ref{sec:flux_comp}) and deposit in cell.
		\EndWhile
	\EndFunction
	\\

	\Function{ray\_fltr}{$p$}
	\State \Return $(p.\ell < \ell_{max}) \land ( p.\tau < \tau_{max} ) \land (p.s < \sqrt{3}L_{\textrm{box}})$
	\EndFunction
	\\

	\Function{s\_copy}{$p_{dst}$, $p_{src}$}
	\Comment{Initialize new rays at copy time}
	\State $(p_{dst}).s \gets (p_{src}).s$
	\State $(p_{dst}).\tau \gets (p_{src}).\tau$
	\State $(p_{dst}).\ell \gets (p_{src}).\ell + 1$
	\Comment{Increment \textsc{HEALPix} level}
	\State $(p_{dst}).L \gets (p_{src}).L/4$
	\Comment{Geometric dilution of luminosity}
	\State $(p_{dst}).i \gets $ \textsc{HEALPix} child index
	\State Assign position, direction to $p_{dst}$ based on $(p_{dst}).i$
	\EndFunction
	\\

	\Function{s\_tsfm}{$p$}
	\Comment{Transform parent ray into child}
	\State \Call{s\_copy}{$p$,$p$}
	\EndFunction

\end{algorithmic}
\caption{User-defined functions, which are inputs to higher-order functions in Alg.~\ref{alg:trace_rays}. Let $p$ denote a single ray. See Sec.~\ref{sec:user_defined} for definitions of other symbols.}\label{alg:functors}
\end{algorithm}
We define our own functions as inputs to the higher-order \textsc{AMReX} primitives in the previous section.
For pseudocode, see Alg.~\ref{alg:functors}.
\par
The \textsc{raymarch} function takes an input ray $p$ and performs ray marching, advancing the ray one cell at a time using a fast voxel traversal technique~\citep{Amanatides1987}. At each cell, flux is computed (Sec.~\ref{sec:flux_comp}) and deposited into both the cell and the neighbor cell closest to the ray, making use of a geometric correction (Sec.~\ref{sec:noise_fix_method}).
\par
Marching continues until one of three stopping conditions is satisfied:
\begin{enumerate}
	\item The splitting criterion (Eq.~\ref{eq:split_criterion}) is satisfied.
	\item $p$ exits its local subdomain box $D_L$.
	\item Distance traversed $p.s$ exceeds $\sqrt{3}L_{\textrm{box}}$. (Periodic BC case.)
\end{enumerate}
\par
The \textsc{ray\_fltr} function takes a ray as input and returns \texttt{True} if its \textsc{HEALPix} refinement level $\ell$, optical depth, and total travel distance are below prescribed thresholds, and \texttt{False} otherwise.
\par
The \textsc{s\_copy} function is used to initialize each of the newly-created rays after splitting.
From the parent, $p_{src}$, the travel distance $s$ and optical depth $\tau$ are copied to the child $p_{dst}$.
Given that the parent is a \textsc{HEALPix} sample at refinement level $\ell$, the children are at refinement level $\ell + 1$.
The original luminosity of the source $L$ is divided evenly among each child, allowing us to account for geometric dilution without an explicit inverse square term.\footnote{In our actual implementation, instead of $V_{shell}$ in the denominator of Eq.~\ref{eq:lut_eqn}, we use $V_{box} = h^3$, where $h$ is the cell width.}
The position of each child can be computed using the distance traveled $s$ and \textsc{HEALPix} child index $i$.
The direction is set by normalizing the ray's position vector relative to the source.
\par
\textsc{s\_tsfm} is called on the parent ray as part of the splitting process. Within this function, the parent ray calls \textsc{s\_copy} on itself, transforming the parent ray into one of its own children. This allows us to avoid allocation of unnecessary space and a subsequent delete operation.

\subsubsection{Ray tracing loop}\label{sec:rt_loop}
\begin{algorithm}
\begin{algorithmic}
\State \Call{spawn\_rays}{}()
\Comment{Init. rays at sources}
\Repeat
\State \Call{transformParticles}{\textproc{raymarch}}
\State \Call{filterParticles}{\textproc{ray\_fltr}}
\State \CallTempl{filterCopyTransformParticles}{3}{\textproc{s\_copy, need\_split, s\_tsfm}}
\State \Call{Redistribute}{}()
\Until {$|P| = 0$}
\Comment{Repeat until set of rays is empty}
\end{algorithmic}
\caption{Ray tracing loop (Sec~\ref{sec:rt_loop}). User-defined functions are given in Alg.~\ref{alg:functors}. $P$ denotes the set of rays. } \label{alg:trace_rays}
\end{algorithm}

Putting it all together, we arrive at our ray tracing loop. Pseudocode is given in Alg.~\ref{alg:trace_rays}. See Figs. \ref{fig:raysplit} and \ref{fig:boundary} for illustration.
First, \textsc{spawn\_rays} creates, at each source, a starting set of rays, at the 0th \textsc{HEALPix level}. Next, \textsc{TransformParticles} is used to march rays forward until stopping criteria are met.
Ray termination is implemented with \textsc{ray\_fltr}, and ray splitting with \textsc{filterCopyTransformParticles}. \textsc{Redistribute} checks if each ray is still in the subdomain owned by the current MPI rank.
If not, parallel communication is performed and the ray is sent to the correct MPI rank. If periodic BCs are in effect (Sec.~\ref{sec:adaptive_periodic}), \textsc{Redistribute}  ensures that rays ``wrap around'' correctly.
The loop repeats until there are no rays left to process.

\subsection{Geometric correction}\label{sec:noise_fix_method}
\subsubsection{Background}
Recall from Section~\ref{sec:geom_correction_literature} the geometric correction proposed by~\cite{wise_enzomoray_2011}.
Specifically, it accounts for the overlap between the volume swept by the ray and the ``current'' cell $c$, multiplying incident flux $\Gamma$ with a correction factor $f_c \in [0, 1]$.
Consider a ray segment passing through cell $c$.
Let $L_{pix}$ denote the linear width of a HEALPix pixel, and $D_{\mathrm{edge}}$ the perpendicular distance between the midpoint of the segment and the nearest cell edge.
(See Figs.~1 and 2 in \cite{wise_enzomoray_2011} and \cite{hartley_arc_2019}, respectively.)
$f_c$ is defined as $(1/2 + D_{edge}/L_{pix})^2$ by~\cite{wise_enzomoray_2011}, but we use the definition from \cite{hartley_arc_2019} instead:
\begin{equation}\label{eq:correction}
	f_c=\begin{cases}
		\frac{1}{2} + \frac{D_{edge}}{L_{pix}} & \text{if $D_{edge} < \frac{L_{pix}}{2}$},\\
		1& \text{otherwise}
	\end{cases}
\end{equation}
\cite{wise_enzomoray_2011} claim that photon loss when $f_c < 1$ is negligible, however \cite{hartley_arc_2019} believe that losses are significant enough to justify a secondary correction, which deposits any flux that falls outside of $c$ into the ``nearest'' neighboring cell $c'$:
\begin{equation}\label{eq:secondary_correction}
	f'_c = (1 - f_c)\frac{a_{abs}}{a'_{abs}}
\end{equation}
where $a_{abs}$ is the absorber (neutral hydrogen) density in the current cell, $a'_{abs}$ is the absorber density in the neighbor, and  $f'_c$ is the correction factor for the \emph{neighboring} cell, i.e. $f'_c \cdot \Gamma$ is deposited into $c'$.
\par
We verify the results of \cite{hartley_arc_2019} with our independent ray tracing implementation, which uses a different time integrator.
In Fig.~\ref{fig:my_fig12_repro}, we attempt to reproduce Fig. 12 of \cite{wise_enzomoray_2011}, which compares Eqs.~\ref{eq:correction} and \ref{eq:secondary_correction} to the formulation in~\cite{wise_enzomoray_2011} on an analytically tractable problem.
Our results match closely, confirming that the secondary correction has a significant effect, and that squaring $f_c$ can increase photon loss if conservation is not enforced.
\par
Note that the absorber density is not stored explicitly on our grids. Instead, we store the hydrogen number density $n_H$ and the ionization fraction $x$ on our grid, then compute $a_{abs} = (1 - x)n_H$.
\subsubsection{Numerical instability for low $n_H$ neighbor cells}
\begin{figure*}
	\centering
	\begin{subfigure}[b]{0.32\textwidth}
		\includegraphics[width=\textwidth]{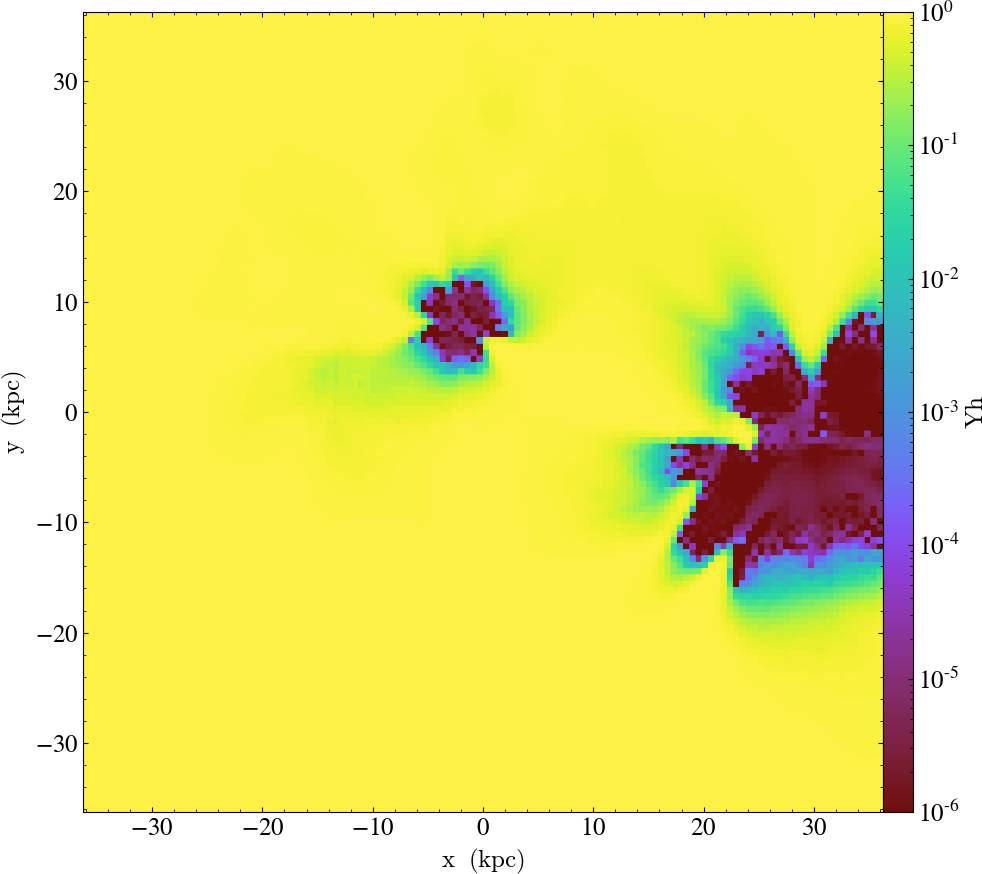}
		\caption{Current SOTA}\label{fig:naive_test4}
	\end{subfigure}
	~
	\begin{subfigure}[b]{0.32\textwidth}
		\includegraphics[width=\textwidth]{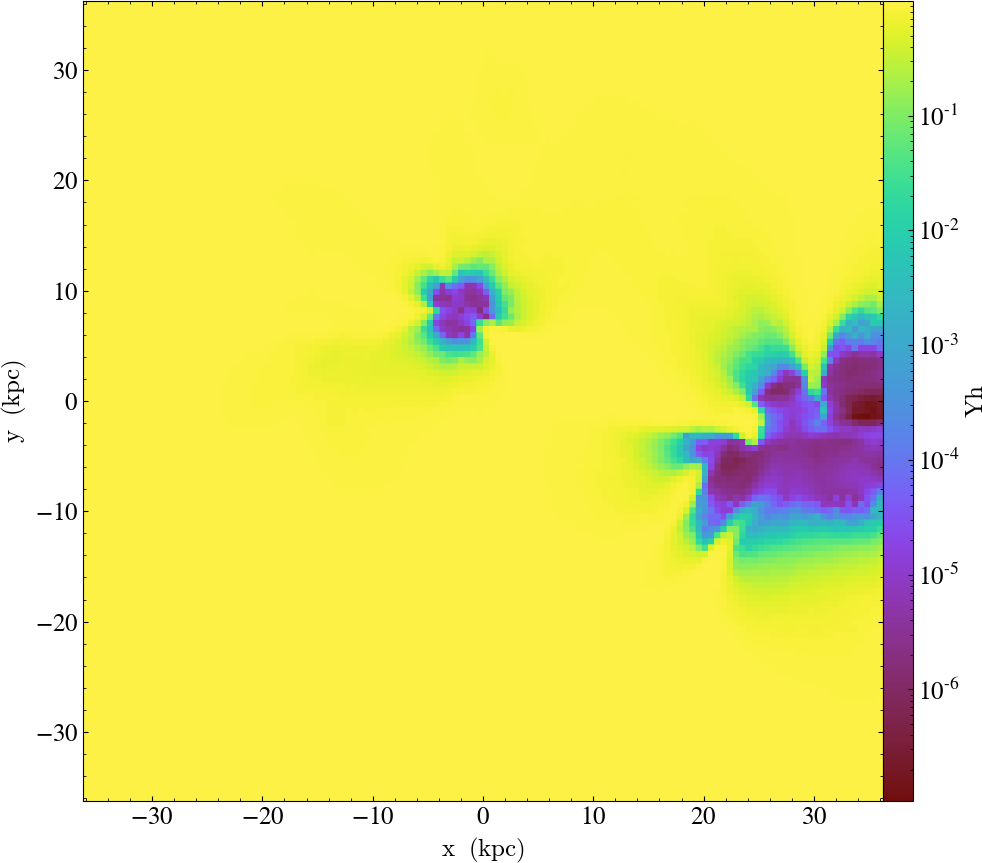}
		\caption{Naive thresholding 1}\label{fig:naive_thresh4}
	\end{subfigure}
	~
	\begin{subfigure}[b]{0.32\textwidth}
		\includegraphics[width=\textwidth]{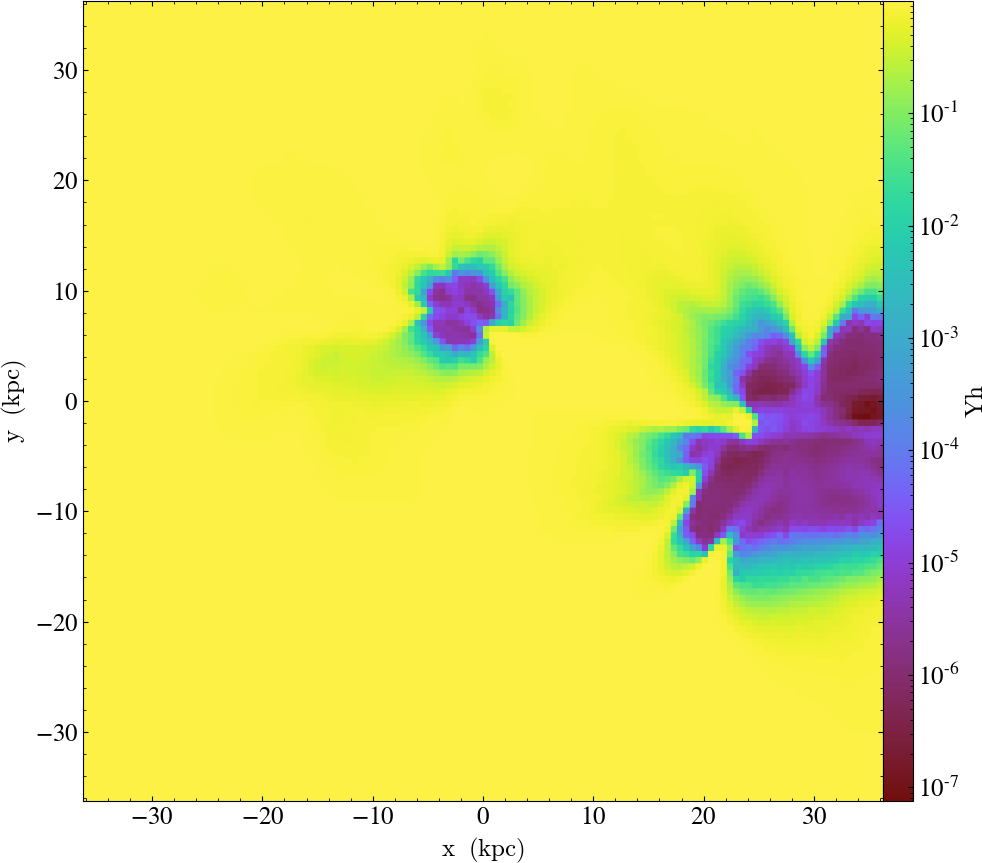}
		\caption{Conservative thresholding}\label{fig:cnsrv_thresh4}
	\end{subfigure}
	\caption{Comparison of ``Test 4'' results (Sec.~\ref{sec:realisticIC}) given
		different solutions to the low $n_H$ edge case described in
	Sec.~\ref{sec:noise_fix_method}.}\label{fig:noise_test4}
\end{figure*}

\begin{figure}
	\centering
	\begin{subfigure}[b]{0.23\textwidth}
		\includegraphics[width=\textwidth]{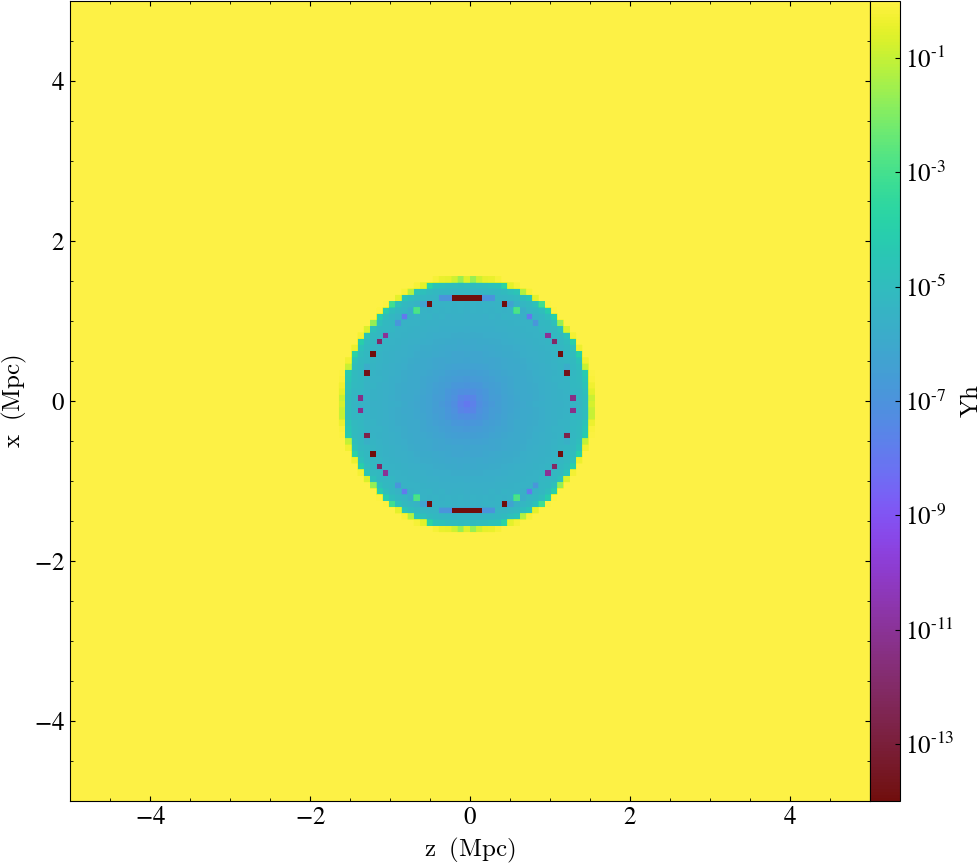}
		\caption{Current SOTA}\label{fig:strom_naive}
	\end{subfigure}
	~ 
	\begin{subfigure}[b]{0.23\textwidth}
		\includegraphics[width=\textwidth]{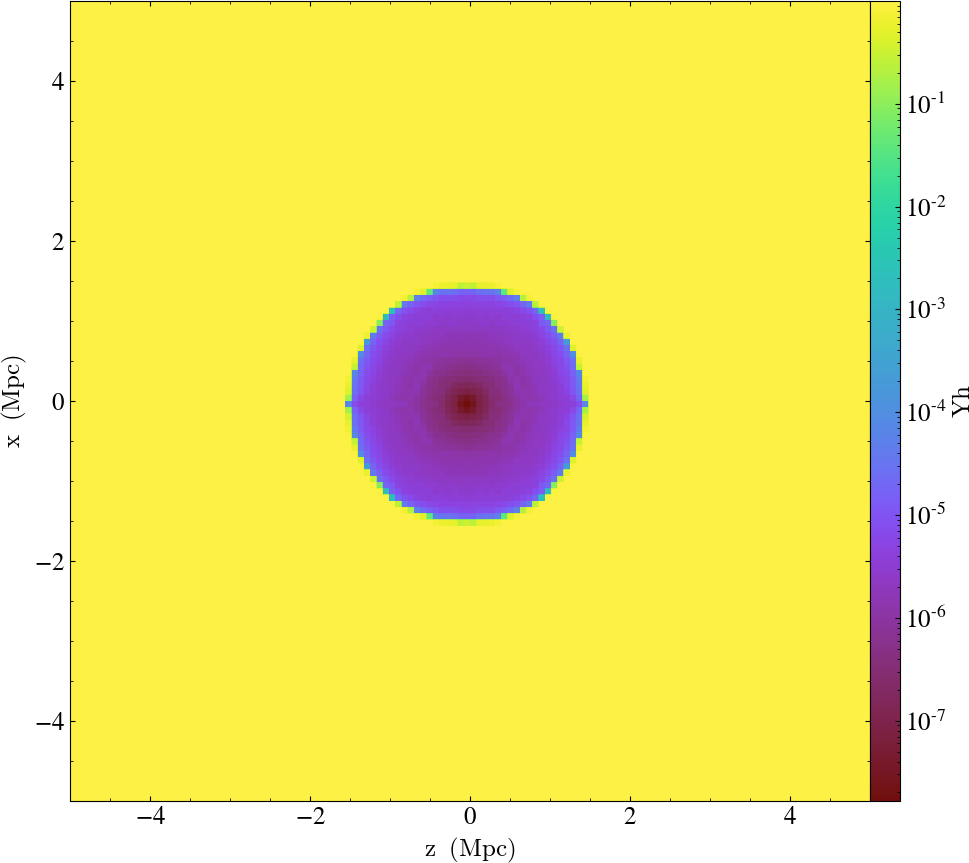}
		\caption{Naive thresholding 1}\label{fig:strom_naive_threshA}
	\end{subfigure}\\
	\begin{subfigure}[b]{0.23\textwidth}
		\includegraphics[width=\textwidth]{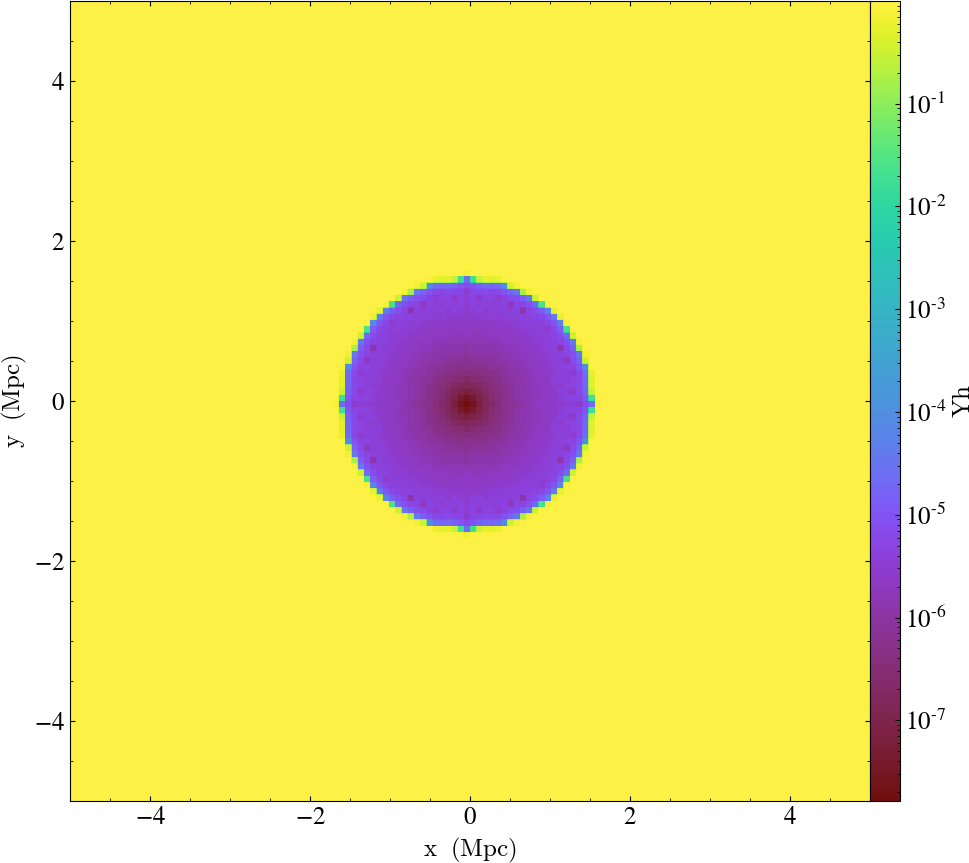}
		\caption{Naive thresholding 2}\label{fig:strom_naive_threshB}
	\end{subfigure}
	~
	\begin{subfigure}[b]{0.23\textwidth}
		\includegraphics[width=\textwidth]{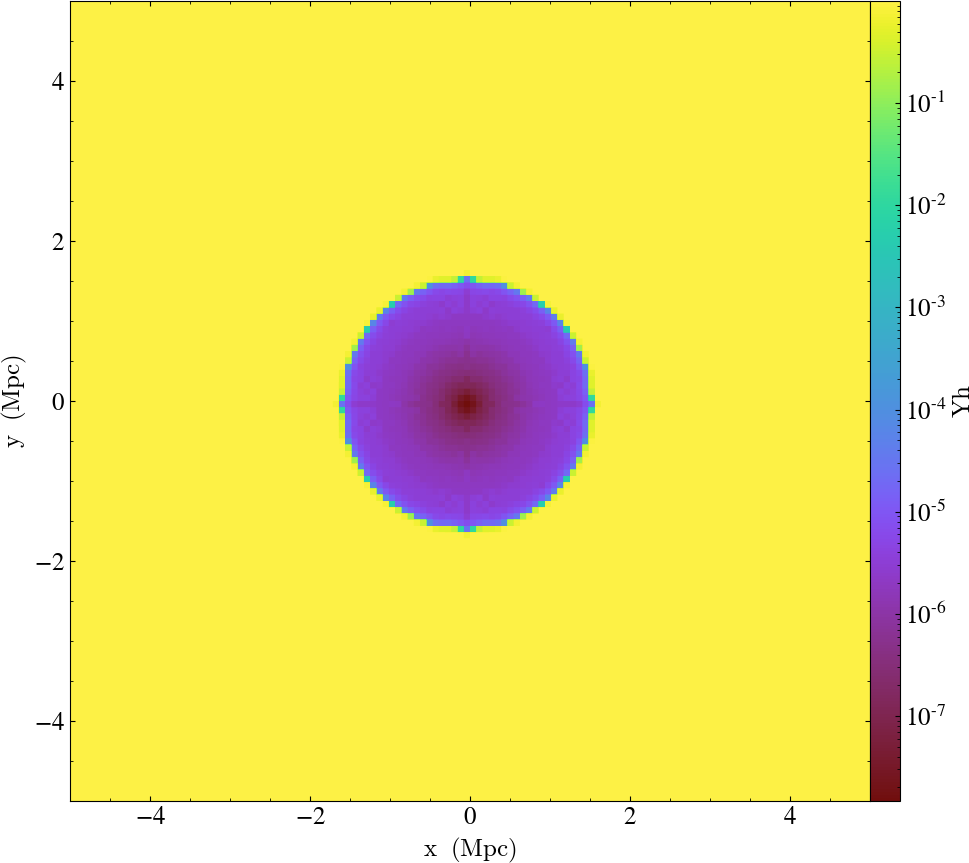}
		\caption{Conservative thresh.}\label{fig:strom_consv_thresh}
	\end{subfigure}
	\caption{Different approaches to handling the low $n_H$ edge case
		(Sec.~\ref{sec:noise_fix_method}), illustrated with a Stromgren sphere
	test.}\label{fig:noise_stromgren}
\end{figure}

Although the secondary correction (Eq.~\ref{eq:secondary_correction}) allows us to conserve photons, it can introduce numerical difficulties.
Small values of $n'_H$ can cause single-cell discontinuities (See Figs.~\ref{fig:strom_naive}, \ref{fig:our_test4}, and \ref{fig:naive_test4}) due to a near-zero denominator in Eq.~\ref{eq:secondary_correction}.
\par
We first attempted to set $f'_c$ to zero if the value of $n'_H$ falls below a given threshold. In other words, if $n'_H$ is small, drop the secondary correction. The advantage of this solution is that it retains the primary correction. We initially assumed that the number of photons lost would be negligible given the relatively low number of affected cells, and because low-density cells are optically thin.\footnote{For low $n_H$ cells, we also adopt the approach of \cTwoRay, which takes the limit as $n_H \rightarrow 0$ to compute a separate LUT for the optically thin case. However, this is not applicable to the secondary correction in Eq.~\ref{eq:secondary_correction}.} This successfully removed the numerical artifacts but visibly reduced the size of the ionized region (Fig.~\ref{fig:strom_naive_threshA},~\ref{fig:naive_thresh4}), demonstrating that photon losses were in fact non-negligible.
\par
Tightening the threshold appeared to mitigate most of these losses. However, it brought back some of the numerical artifacts, which, upon close inspection, are still slightly visible in Fig.~\ref{fig:strom_naive_threshB}.
\par
A more effective solution is to go back to the higher threshold, while enforcing conservation: if we skip depositing flux into the neighbor, all of it goes back into the original cell. We modify Eqns.~\ref{eq:correction} and~\ref{eq:secondary_correction}:
\begin{align}\label{eq:new_correction}
	f_c &= \begin{cases}
		\frac{1}{2} + \frac{D_{edge}}{L_{pix}} & \text{if $(D_{edge} < \frac{L_{pix}}{2}) \land (n'_H \geq n_{th})$},\\
		1& \text{otherwise}
	\end{cases}\\
	f'_c &= \begin{cases}\label{eq:new_secondary}
		(1 - f_c)\frac{(1-x)n_H}{(1-x')n'_H} & \text{if $n'_H \geq n_{th}$},\\
		0 & \text{otherwise}
	\end{cases}
\end{align}

This approach is not ideal because we lose the original geometric correction (Eq.~\ref{eq:correction}) in the low $n_H$ case.
However, we find that this is a reasonable tradeoff: in Figs.~\ref{fig:strom_consv_thresh} and~\ref{fig:cnsrv_thresh4}, it can be seen that this ``conservative thresholding'' allows us to avoid both numerical artifacts and photon loss.

\subsection{Coupled simulation pipeline}\label{sec:coupled_method}

\begin{figure*}
	\begin{center}
		\includegraphics[width=0.95\textwidth]{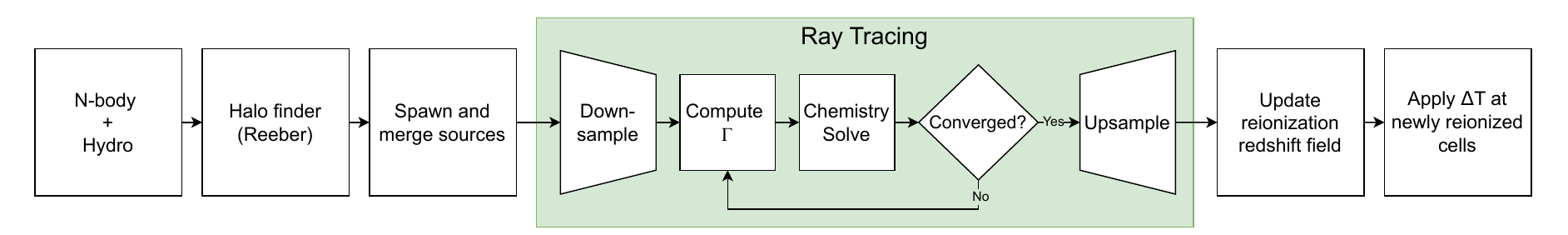}
	\end{center}
	\caption{Illustration of a single timestep in our coupled simulation pipeline. (Sec~\ref{sec:coupled_method}.)}\label{fig:flowchart}
\end{figure*}

In order to couple ray tracing to the rest of the Nyx simulation, specifically $N$-body, hydrodynamics, and radiative heating / cooling, we perform the following at each timestep:
\begin{enumerate}
	\item Localize point sources.
	\item Ray trace on a downsampled density field.
	\item Account for photoheating.
\end{enumerate}
See Fig.~\ref{fig:flowchart} for an illustration of this pipeline. In the following subsection, we will describe each of these steps in detail.
\subsubsection{Halo finding and source luminosity}\label{sec:point_sources}
We assume that ionizing radiation comes only from galactic halos, which act as point sources.
After each timestep, the Reeber library is called to construct a distributed merge tree~\citep{Morozov_2013} representing the topology of the density field.
The merge tree is used to localize halos and find their masses per the method of~\cite{Friesen2016}.
\par
To find the number of ionizing photons emitted per unit time $\dot N_\gamma$ from each halo, we assume a linear relationship between halo mass and ionizing flux, where $M_h$ is the halo mass and $\zeta$ is a constant:
\begin{equation}
	\dot N_\gamma = \zeta M_h
	\label{eq:mass_to_light}
\end{equation}
We denote $\zeta$ as the \emph{photon conversion ratio} or \emph{mass-to-light constant}. See Sec.~\ref{sec:large_scale_simulation} for how we estimate a lower bound on the value of $\zeta$.
\subsubsection{Radiation step}\label{sec:rad}
After each $N$-body and hydrodynamics timestep of length $\Delta t$, the density field is downsampled onto a coarse grid for ray tracing.
The radiative transfer solution, on the coarse grid, is advanced by $\Delta t$.
The ionization fraction field is then upsampled to the fine grid.
Next, for each newly-reionized cell (ionization fraction exceeds 90\% for the first time), the current redshift (i.e. the time of reionization) is recorded to a channel that we denote as the \emph{reionization redshift field}. 
\subsubsection{Heating due to reionization}\label{sec:heating}
\par
To account for heating due to reionization, we modify the approximation method used by~\cite{onorbe_inhomogeneous_2019}. We read the reionization redshift field (Sec.~\ref{sec:rad}) and apply a fixed temperature jump $\Delta T$ at the time of reionization. $\Delta T$ is a free parameter, and in this paper we choose $\Delta T = \num{2.0e4}.$
\par
The key difference between our method and~\cite{onorbe_inhomogeneous_2019} is that they precompute the reionization redshift field using a semi-numerical method, whereas we update it on-the-fly based on ray tracing output.

\subsection{Source Merging Algorithm}\label{sec:merging}
After reionization begins, the number of radiating sources can grow rapidly, which presents performance and scalability challenges: computational cost is roughly linear in the number of sources. Depending on the resolution and halo mass threshold, the number of sources at $z \sim 6$ can reach into the hundreds or thousands. Furthermore, computational costs compound when ionized bubbles expand due to their optically thin interiors; as I-fronts propagate there are fewer opportunities for ray termination from optical depth (Sec.~\ref{sec:tracing}).
\par
To mitigate the computational cost of ray tracing, various techniques to decrease either the number of sources or number of raycasts per source have been developed.
Recall that adaptive ray tracing (Sec.~\ref{sec:tracing}) is one way to address the large number of redundant ray traversals found in long characteristics methods (Sec.~\ref{sec:past_short_long_char}).
\emph{Ray merging} is also discussed in the literature: \cite{abel_adaptive_2002} recombine child rays (and their fluxes) into a single ray traveling in the parent direction when angular resolution requirements decrease. \cite{Trac_2007} introduce a ray merging scheme that enforces an upper bound on the number of active rays, and \cite{Cain_2024} is a recent work that implements this scheme.
\par
It is also possible to reduce the effective number of sources using a tree-based source aggregation or `source merging' approach~\citep{Razoumov2002,Hasegawa2010,Okamoto2012,wunsch_tree-based_2018, grond_trevr_2019}.
In these works, rays are not cast from every point source individually.
Instead, all sources are inserted into a spatial tree; when rays are traced backward from a cell (or target region), distant sources that subtend a sufficiently small opening angle at the target are grouped into a single effective emitter node in the tree.
This takes the source dependence of the algorithm from linear to logarithmic (as the depth of the tree).
However, the backwards ray tracing assumption appears to be fundamentally at odds with our use of adaptive ray tracing, which is a \emph{forward} algorithm.
Therefore, it is unclear how to map this algorithmic improvement to our own.
\par
Inspired by the source merging processes in \cite{grond_trevr_2019, wunsch_tree-based_2018}, we develop a novel forward source merging process that groups sources within the same ionized bubble.
This is physically motivated by the assumption that within a large, ionized (and optically thin) bubble, photons emitted from multiple sources traverse similar paths, all of which are low opacity.
Therefore, any given distant neutral region will receive their effective combined radiation as if it came from a single emitter of equivalent luminosity.
The two algorithms in this process are outlined in Alg.~\ref{alg:source_merge}. In the following paragraphs we will describe determination of bubble radii around individual sources, radius comparison, and collapsing individual sources within the same bubble.
\par
We first collect the per-rank local source lists on the designated I/O processor.
There, duplicates (sources assigned to the same grid cell) are identified and merged by summing their fluxes.
The resulting deduplicated list is then sorted and broadcast back to all ranks, so that each MPI rank operates on a unified, globally ordered source catalog.
This synchronization is essential before computing bubble radii and applying source merging, since it ensures that every rank has a consistent view of the complete set of emitters.
The merging behavior is globally controlled by a user-specified `automerge' parameter: once the global average ionized fraction surpasses the automerge threshold, the algorithm merges sources in all subsequent timesteps.
\par
For each source, we must compute the radius of the surrounding ionized bubble to determine whether another source is within that bubble.
To achieve this, we sample the ionization field on concentric spherical shells.
We define a shell of trial radius $R$ centered at the source, initialized to zero but to be incremented by $\Delta r$ in code units.
Assuming a uniform grid, our default $\Delta r = 2$ corresponds to a comoving distance of $\Delta r = 2{L_{box}}/{N_{res}}$ where $L_{box}$ is the length of one volume dimension and $N_{res}$ is the resolution of the box in number of cells.
\par
Over a sphere of radius $R$, we sample the ionization fraction field at a fixed angular resolution. Let $x(\mathbf{i})$ denote the ionization fraction stored on the grid at cell index $\mathbf{i}$.
A sampled cell is counted as ionized if $x(\mathbf{i}) \ge x_{\rm th}$, where $x_{\rm th}\in[0,1]$ is a user-specified threshold.
We evaluate $x(\mathbf{i})$ at $N_{\rm samp}$ approximately uniform angular samples on the shell, defining
$N_{\rm tot}(R)$ as the number of evaluated samples and $N_{\rm ion}(R)$ as the number satisfying $x(\mathbf{i})\ge x_{\rm th}$.
The shell is passed if $N_{\rm ion}(R)/N_{\rm tot}(R) \ge f_{\rm req}$, where $f_{\rm req}\in(0,1]$ is threshold on ionized fraction.
When $N_{\rm tot}(R) < N_{\min}$, where $N_{\min}$ is a user-specified threshold, we require all evaluated samples to satisfy $x(\mathbf{i})\ge x_{\rm th}$ in order for the shell to pass.

If a shell passes, we continue our search and increase the radius by $\Delta r$. Otherwise we stop and return the current value of $R-\Delta r$ as the bubble radius of that source.
We also impose a hard upper limit $R_{max}$ on shell expansion to prevent over-merging once most of the domain is ionized.
This is the most sensitive user-defined parameter.
A larger $R_{max}$ allows bubbles to span greater distances, increasing the chance of overlap and making source merging merging sources more aggressive. A smaller $R_{max}$ curtails bubble growth, reduces overlap, and permits finer source granularity.
\par
Once all bubble radii are determined, the merge algorithm identifies groups of sources whose bubble radii overlap, compares their distances, and then uses a union–find structure to find connected components.
For two example sources $a$ and $b$, with bubble radii $r_a$ and $r_b$ and coordinates $\mathbf{x}_a$ and $\mathbf{x}_b$ computed under minimum-image (periodic) geometry, the squared distance between them is given as:
\begin{equation*}
d^2 = \lVert \mathbf{x}_a - \mathbf{x}_b \rVert^2 \quad
\end{equation*}
This distance is compared to a combined overlap radius set via a user-defined overlap factor $f_{\rm ov}$:
\begin{equation*}
r_{\rm sum} = f_{\rm ov} \bigl(r_a + r_b\bigr).
\end{equation*}
The sources are considered overlapping if:
\begin{equation*}
 d^2 \le r_{\rm sum}^2 + \epsilon.
\end{equation*}
where $\epsilon$ is a small tolerance to guard against floating-point error.
Pairwise overlaps such as these define edges in a graph whose connected components are computed via a union–find data structure, yielding groups of mutually overlapping sources. Each group is then collapsed into a single merged source with the merged flux being the sum of member fluxes.
The merged radius is the maximum among member radii and the position is chosen by computing the flux-weighted centroid and then snapping to the nearest original source position.
The final merged catalog is returned in a deterministic sorted order, which is read rank-by-rank into the ray tracing algorithm.
\par
\begin{figure}
    \centering
    \includegraphics[width=0.9\linewidth]{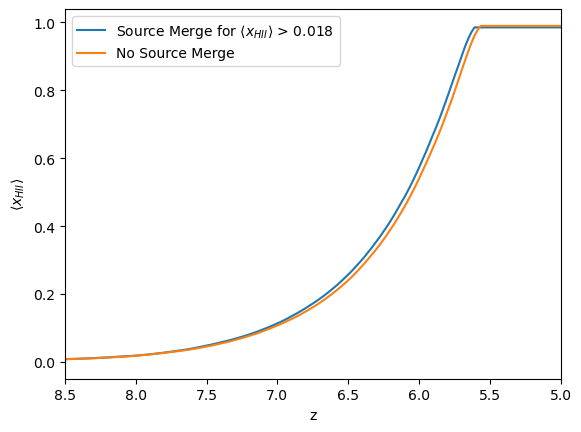}
    \caption{Comparison of ionized fraction as a function of redshift between ray tracing with (blue) and without (orange) source merging for $1024^3$ coupled simulations. }
    \label{fig:sm_xh}
\end{figure}

Depending on user-defined parameters, using source merging can decrease the number of sources emitting rays by an arbitrarily large percentage.
However, to achieve qualitatively similar ionization fraction redshift evolution relative to non-merged simulations (see \ref{fig:sm_xh}) , we found the lower limit of post-merged sources to be a factor of three less than the original.
Generally, the greater the number of sources, the more accurate the ionization fraction field.
\par
Figure \ref{fig:sm_xh} demonstrates agreement between $1024^3$ simulation results, with and without source merging. Both simulations down-sample the $1024^3$ by a factor of 4 to a $256^3$ grid for ray tracing.
Merging sources does lead to slightly earlier reionization, due to an overestimation of bubble size caused by source grouping.
Figure \ref{table:slices_sm} plots the ionization fraction field on a slice of the domain for redshifts of $z \sim 7$ and $z\sim 6$.
Although the exact bubble shape shifts slightly, the patchy non-spherical process of reionization and similar average ionized fraction values are maintained.
Across these simulations (at $1024^3$), wall time from $z \sim 199$ to $z \sim 5$ was halved with source merging on.
\begin{algorithm}
\begin{algorithmic}

\Function{max \_radius}{$S$}
    \Comment{$S$ has integer coords $S.\mathbf{i}$}
    \State $x_0 \gets$ ionization fraction at $S.\mathbf{i}$
    \If{$x_0 < x_{\rm th}$}
        \State \Return $0$
    \EndIf

    \State $r \gets \Delta r$
    \While{$r \le R_{\max}$}
        \State $\mathcal{P}(r) \gets$ checked point samples on the shell of radius $r$

         \State $N_{\rm tot}(r)\gets |\mathcal{P}(r)|$
         \State $N_{\rm ion}(r)\gets |\{\mathbf{p}\in\mathcal{P}(r): x(\mathbf{p})\ge x_{\rm th}\}|$

        \If{$N_{\rm tot}(r) < N_{\min}$}
            \If{$\exists\,\mathbf{p}\in\mathcal{P}(r)\ \text{s.t.}\ x(\mathbf{p})<x_{\rm th}$}
                \State \Return $r - \Delta r$
            \EndIf
        \ElsIf{$N_{\rm ion}(r)/N_{\rm tot}(r) < f_{\rm req}$}
            \State \Return $r - \Delta r$
        \EndIf

        \State $r \gets r + \Delta r$
    \EndWhile

    \State \Return $R_{\max}$
\EndFunction
\\

\Function{merge\_sources}{$\{S_k\}$}
    \Comment{$S_k$ is the $k$th source}
    \State Initialize union--find $\mathcal{U}$

    \ForAll{pairs $(a,b)$}
        \State $d \gets$ min-image distance between centers
        \If{$d \le f_{\rm ov}(r_a+r_b)$ and $r_a,r_b>0$}
            \State \Call{unite}{$\mathcal{U},a,b$}
        \EndIf
    \EndFor

    \State Form groups $\{\mathcal{G}_m\}$ from $\mathcal{U}$
    \State $\mathcal{M} \gets \emptyset$
    \Comment{Set of merged sources}

    \ForAll{groups $\mathcal{G}_m$}
        \If{$|\mathcal{G}_m|=1$}
            \State Append single source to $\mathcal{M}$
            \State \textbf{continue}
        \EndIf

        \State $F_{\rm tot} \gets$ sum of fluxes in group
        \State $R_{\rm max} \gets$ largest radius in group

        \State Compute flux-weighted centroid (with periodic unwrapping)
        \State Round centroid and wrap back into domain
        \State Snap to nearest original member cell

        \State Create merged source $M$ with:
        \State \hspace{1em}$M.{\rm flux}=F_{\rm tot}$,\;
                          $M.{\rm radius}=R_{\rm max}$,\;
                          $M.\mathbf{i}=\text{snapped cell}$

        \State Append $M$ to $\mathcal{M}$
    \EndFor

    \State Sort $\mathcal{M}$ lexicographically
    \State \Return $\mathcal{M}$
\EndFunction

\end{algorithmic}
\caption{source-radius and source-merging algorithm.} \label{alg:source_merge}
\end{algorithm}

%

\begin{table*}
\centering
\imagetablefour{
	figures/sourcemerge/x_H00350_x,
	figures/sourcemerge/x_H00390_x,
	figures/sourcemerge/x_H00350_x_sm,
        figures/sourcemerge/x_H00390_x_sm,
}
\captionof{figure}{Slices of ionization fraction field at $1024^3$ resolution, with (right) and without (left) source merging at $z= 6.9$ and $z=5.8$. For the simulation with source merging, merging was automatically applied at avg. $x_{\mathrm{HII}} > 0.018$ and kept on for the duration of the run. }
\label{table:slices_sm}
\end{table*}

\section{Verification and Validation} \label{sec:eval}

\begin{figure*}
	\centering
	\begin{subfigure}{0.26\textwidth}
		\includegraphics[width=\textwidth]{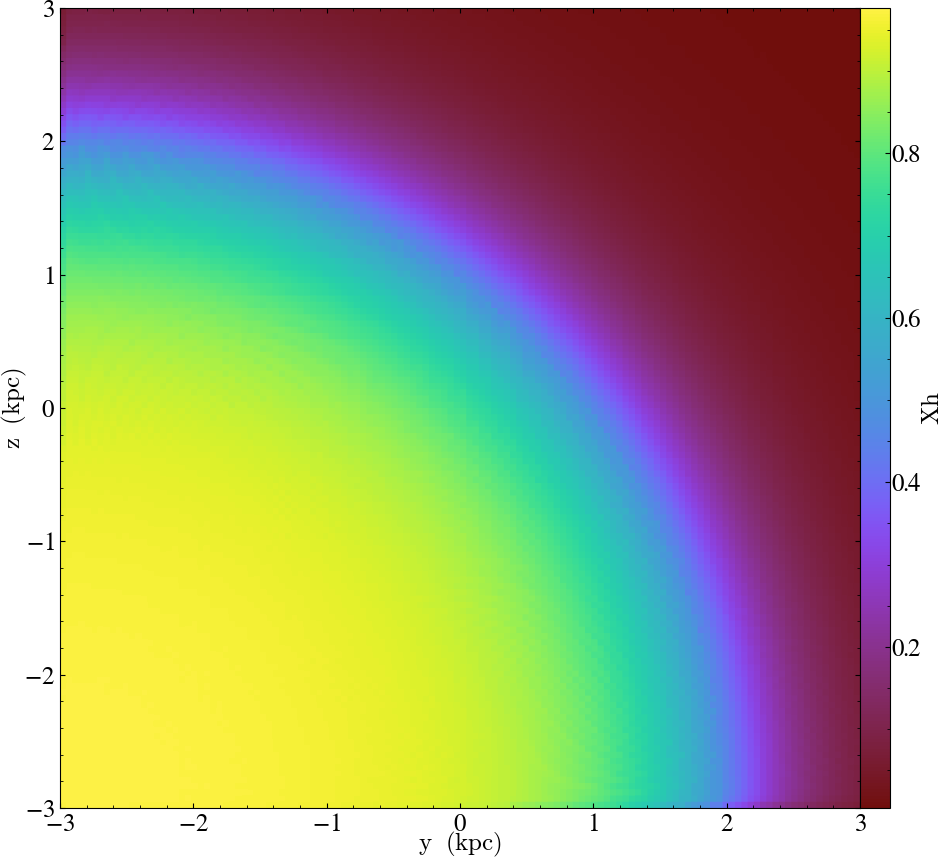}
		\caption{Ionization fraction, Test A}\label{fig:testA}
	\end{subfigure}
	~
	\begin{subfigure}[b]{0.32\textwidth}
		\includegraphics[width=\textwidth]{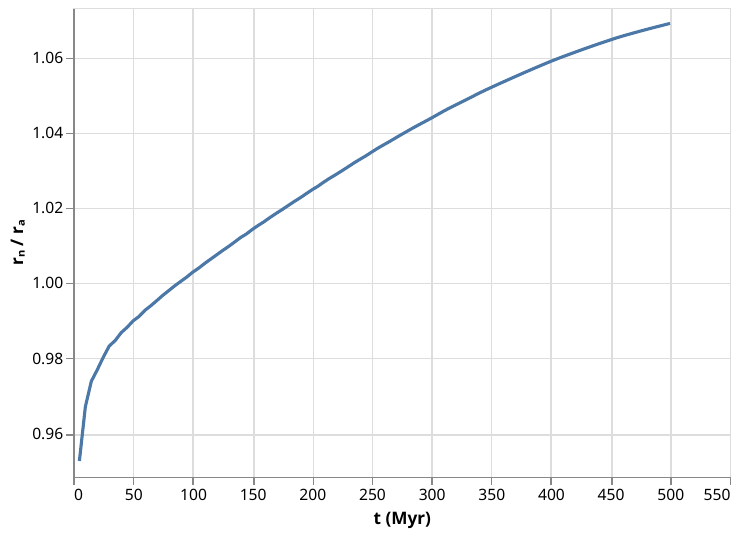}
		\caption{$r_{n}/r_{a}$, Test A}\label{fig:ratio_test2}
	\end{subfigure}
	~
	\begin{subfigure}[b]{0.32\textwidth}
		\includegraphics[width=\textwidth]{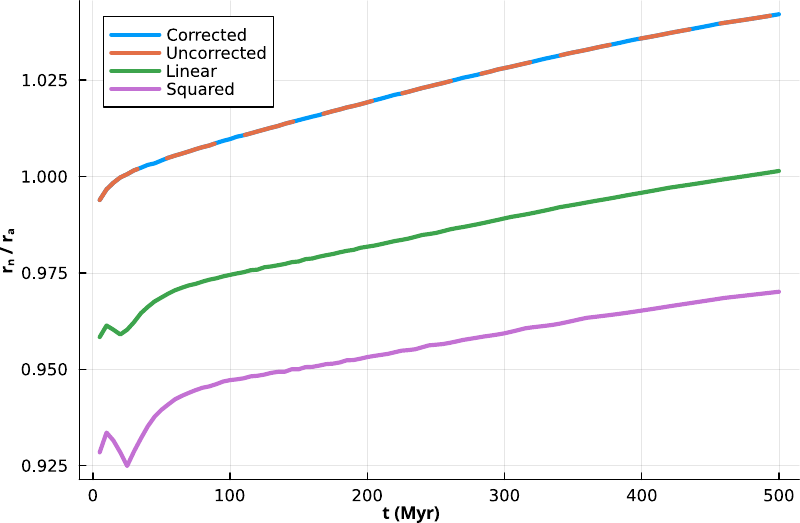}
		\caption{Geometric correction, Test A}\label{fig:my_fig12_repro}
	\end{subfigure}
	\caption{Spherical symmetry test results (Sec.~\ref{sec:stromtests}). In~(\subref{fig:ratio_test2}) and (\subref{fig:my_fig12_repro}), $r_n$ denotes the numerical Stromgren sphere radius from our simulation results, and $r_a$ the analytic prediction.
	In (\subref{fig:my_fig12_repro}), we verify the results in Fig. 12 of \cite{hartley_arc_2019} using our independent ray tracing implementation and choice of time integrator (Sec.~\ref{sec:geom_correction_literature}). ``Uncorrected'' denotes the result without any geometric correction. ``Linear'' uses only the correction in Eq.~\ref{eq:correction}. ``Corrected'' uses both Eq.~\ref{eq:correction} and the conservative secondary correction in Eq.~\ref{eq:secondary_correction}. ``Squared'' uses the correction factor $f_c$ as defined in~\cite{wise_enzomoray_2011}.}\label{fig:analyticSelected}
\end{figure*}

\subsection{Str\"{o}mgren sphere tests}\label{sec:stromtests}
Analytic verification was performed by considering the Str\"{o}mgren sphere problem, which has a well-known analytic solution.
Assuming the initial condition of a point source with ionizing flux $\dot{N}_\gamma$ within a region of neutral hydrogen (HI) with density $n_H$, radiation from the source will ionize the gas around it, forming an HII sphere centered at the source.
Assuming the sphere has a sharp boundary, at photoionization equilibrium it will have radius $r_s$~\citep{stromgren1939}:
\begin{equation}
	r_s = \left[ \frac{3\dot{N}_\gamma}{4\pi\alpha_B(T)Cn^2_H} \right]^{1/3}
\end{equation}
where $C$ is the clumping factor and, making use of the On-The-Spot (OTS) approximation, $\alpha_B(T)$ is the recombination coefficient at temperature $T$.
Furthermore, the following expression can be derived for the temporal evolution of the spherical HII region:
\begin{equation}
	r_I(t) = r_s\left[ 1 - \exp(-t/t_{rec}) \right]^{1/3}
\end{equation}
where recombination time $t_{rec} = \left[ C\alpha_B(T)n_H \right]^{-1}$.

\subsubsection{Test cases}
We choose three sets of problem parameters to evaluate against, which are given in Table~\ref{tab:testABCparams}.
Test A is an eighth symmetry model with the source at one corner of the domain, whereas Tests B and C place the source at the center of the domain.
\par
It can be seen from Figure~\ref{fig:analyticSelected} that our results are in close agreement with analytic solutions. The agreement is not perfect but comparable deviations are exhibited by other codes in the literature, for example \cTwoRay{} and \Arc{}.~\citep{mellema_c2-ray_2006, hartley_arc_2019}.
\begin{table}
	\centering
	\caption{Parameters for test problems described in
	section~\ref{sec:stromtests}.
All assume $T = \SI{1e4}{\kelvin}$, thus $\alpha_B(T) = \num{2.59e-13}~\si{\cm\cubed\per\second}$.}
	\label{tab:testABCparams}
	\begin{tabular}{lllll} 
		\toprule
		Param. 						& Units									& Test A 					& Test B 				& Test C\\
		\midrule
		$n_H$             & \si{\per\cubic\cm} 		& \num{1e-3} 			& \num{1.87e-4} & \num{1.87e-4}\\
		$\luminosity$ 		& \si{\per\second} 			& \num{5e48} 			& \num{1e54} 		& \num{1e57}\\
		$C$ 						  & Unitless							& 1 							& 5 						& 1\\
		$L_{box}$ 				& \si{\mega\parsec}			& 0.0066 					& 1.62 					& 10\\
		\bottomrule
	\end{tabular}
\end{table}

\subsubsection{Evaluation of beam overlap correction}
In Figure~\ref{fig:my_fig12_repro}, we attempt to reproduce Fig.~12 of~\cite{hartley_arc_2019}.
This is a key test of our implementation of \Arc{}'s photon-conserving geometric correction method.
(Sec.
\ref{sec:geom_correction_literature} and Appendix~\ref{sec:noise_fix_method}.) Qualitatively, each of the four curves match the \Arc{} results quite closely despite being an independent implementation with fundamentally different approaches for integration over both time and frequency, which is an encouraging result.

\subsubsection{Realistic initial conditions}\label{sec:realisticIC}
\begin{figure}
	\centering
	\begin{subfigure}[b]{0.23\textwidth}
		\includegraphics[width=\textwidth]{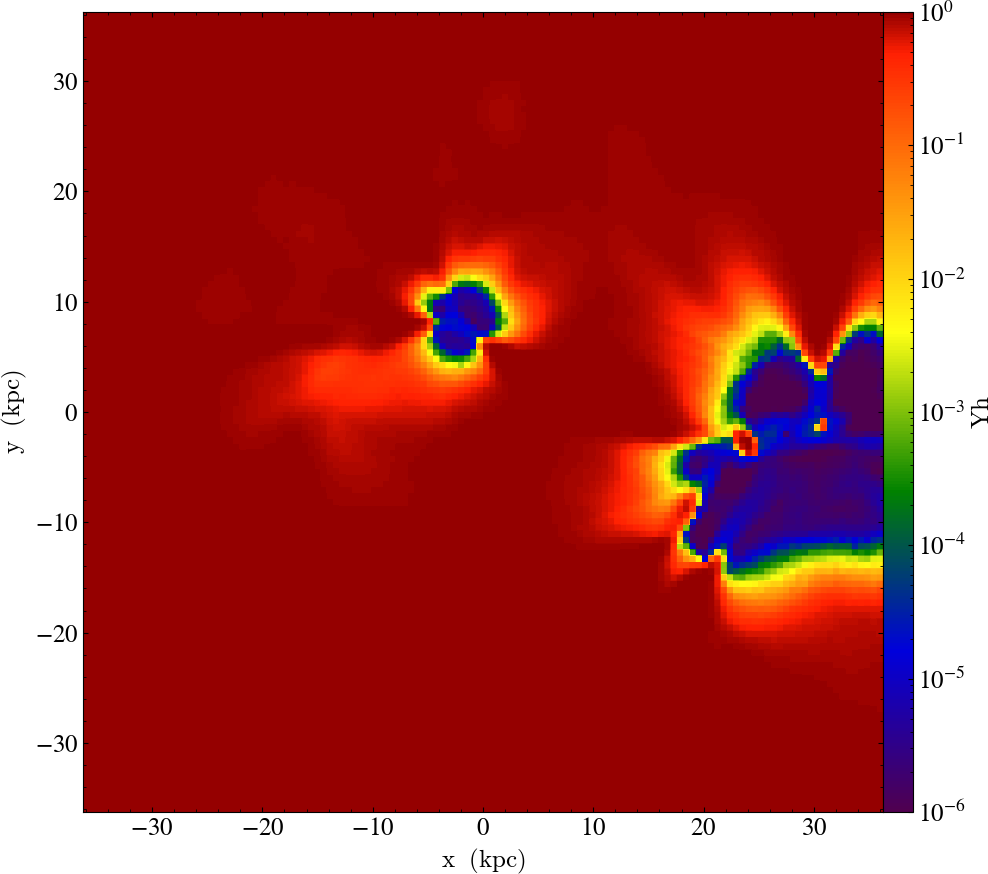}
		\caption{Ours}\label{fig:our_test4}
	\end{subfigure}
	~
	\begin{subfigure}[b]{0.23\textwidth}
		\includegraphics[width=\textwidth]{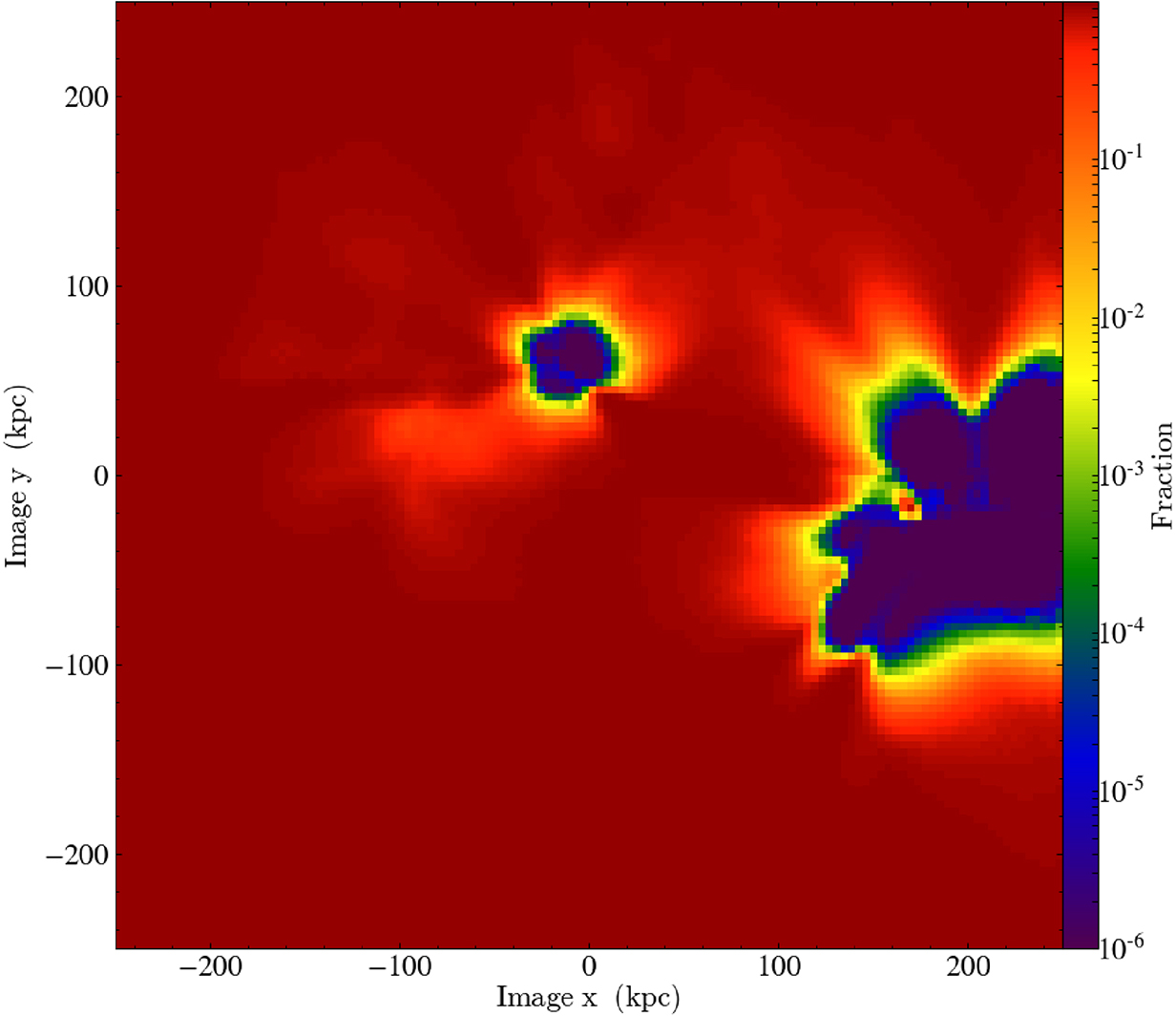}
		\caption{\Arc{}}\label{fig:arc_test4}
	\end{subfigure}
	~
	\caption{Ionization fraction at $t=0.05$ Myr for ``Test 4'' of the
Cosmological Radiative Transfer Comparison Project.
We attempt to match the color map in~(\subref{fig:arc_test4}). See Figure 31 of \cite{iliev_cosmological_2006} for reference. }\label{fig:test4_cmpr}
\end{figure}

In order to go beyond the spherically symmetric geometry of an analytically tractable problem, we tested with the initial density field and source configuration in ``Test 4'' of the Cosmological Radiative Transfer Comparison
Project.~\citep{iliev_cosmological_2006}

Figure~\ref{fig:test4_cmpr} compares a slice of our output to published results from the \textsc{Arc} code.
Qualitatively, the shape and size of our ionized regions appear to be in good agreement with the numerical solutions in the literature.

\section{Large-scale simulation}\label{sec:large_scale_simulation}
\begin{table*}
\imagetable{
	figures/conv_study/N4096s4L40_45695970/density00347_x,
	figures/conv_study/N4096s4L40_45695970/x_H00347_x,
	figures/conv_study/N4096s4L40_45695970/Temp00347_x,
	figures/conv_study/N4096s2L40_45710038/density00449_x,
	figures/conv_study/N4096s2L40_45710038/x_H00449_x,
	figures/conv_study/N4096s2L40_45710038/Temp00449_x,
	figures/conv_study/N4096s1L40_45996344/density00753_x,
	figures/conv_study/N4096s1L40_45996344/x_H00753_x,
	figures/conv_study/N4096s1L40_45996344/Temp00753_x,
}

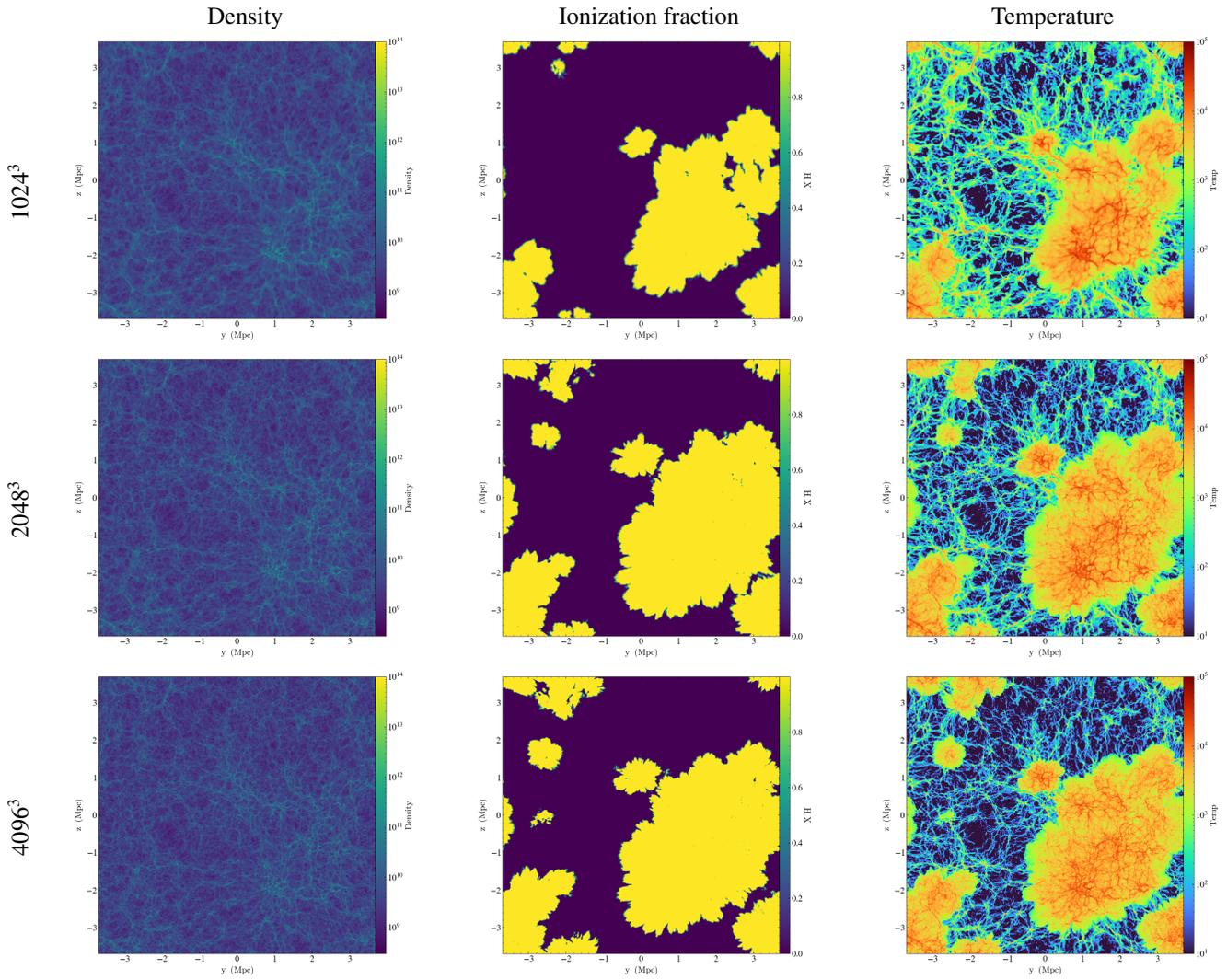
\captionof{figure}[h]{Slices of density, ionization fraction, and temperature fields at time $z=7$ and position $x=0$ at $1024^3$, $2048^3$, and $4096^3$ resolutions in our coupled simulation runs. Axes are in physical coordinates. See Sec.~\ref{sec:sim_results} for more details.}
\label{table:slices7}
\end{table*}

We demonstrate coupled ray tracing (Sec~\ref{sec:coupled_method}) in Nyx for a realistic problem at large scale and at multiple resolutions. In this section, we describe our initial conditions and choice of physical and computational parameters. In addition, we measure computational performance, visualize simulation fields, and plot summary statistics.
\subsection{Simulation parameters}
Simulations were run at resolutions of $1024^3$, $2048^3$, and $4096^3$ on 16, 128, and 1024 nodes of the Perlmutter supercomputer, respectively, where each node contains 4 GPUs.
Ray tracing was performed on downsampled grids of size $256^3$, $512^3$, and $1024^3$.
In order to strike a balance between performance and accuracy, the ray splitting rate \safetyF was set to 2.

\par
We set cosmological parameters to: $\Omega_m = 0.308496, \Omega_b = 0.0488911, h = 0.67742, \sigma_8 = 0.82, n_s = 0.965$, and we initialize the matter density field within a $40~h^{-1}~{\rm Mpc}$ box at $z_i=199$ using $4096^3$ dark matter particles with $6.5\times 10^5~h^{-1}~M_\odot$ each.
The initial conditions for the gas are also initialized on a $4096^3$ mesh, assuming that baryons trace dark matter. We downsample these initial conditions for our $1024^3$ and $2048^3$ runs, to preserve the same realization of modes.
The boundary conditions are triply periodic, which applies to ray tracing (Sec.~\ref{sec:adaptive_periodic}) in addition to hydrodynamics and $N$-body.
\par
A lower bound on the mass-to-light constant $\zeta$ is computed as follows.
Assume a priori that reionization ends at $z \approx 6$. Ignoring recombination, the total number of ionizing photons emitted from the beginning of the simulation $t_0$ until $z=6$ can be computed by summing the luminosity $\luminosity$ of each halo and integrating w.r.t.
time.
This can be expressed in terms of $\zeta$ and halo mass $M_h$ by substituting Eq.~\ref{eq:mass_to_light}:

\begin{equation}
 	\int_{t_0}^{t_{z=6}} \sum_i \luminosity^{(i)} \odif{t} = \int_{t_0}^{t_{z=6}} \sum_i \zeta M_h^{(i)} \odif{t}
	\label{eq:mass_light_lhs}
\end{equation}

Let $\rho_b$ denote the universe's baryonic matter density, $\chi_H$ the H fraction, and $m_p$ the mass of a proton.
Recall that $\rho_b/\rho_c = \Omega_b$, where $\Omega_b$ is the baryon density parameter and $\rho_c = 3H^2/8 \pi G$ is the critical density.
Then the H number density $n_H$ can be computed:

\begin{equation}
	n_H = \frac{\chi_H \rho_b}{m_p} = \frac{\chi_H \Omega_b 3H^2}{m_p 8 \pi G}
	\label{eq:H_number_density}
\end{equation}

Next, let $\Vbox$ denote the volume of the domain.
Then $n_H \Vbox$ is the total number of H atoms.
Assuming hydrogen reionization only, in order for the domain to reionize, the number of photons emitted must be at least the number of H atoms.
In practice, due to recombination, the number of emitted photons must be greater than this.
Therefore, $n_H \Vbox$ bounds the expression in Eq.~\ref{eq:mass_light_lhs} from below. Rearranging, we obtain:

\begin{equation}
	\zeta > \frac{n_H \Vbox}{\int_{t_0}^{t_{z=6}} \sum_i M_h^{(i)} \odif{t}}
	\label{eq:zeta}
\end{equation}

To estimate $\zeta$, we disabled ray tracing and ran Nyx at $4096^3$ with halo finding (Sec.~\ref{sec:point_sources}) enabled in order to obtain halo masses $M_h^{(i)}$ at each timestep with masses above $10^{10}/h~\unit{\solarmass}$. We numerically integrated the denominator of Eq.~\ref{eq:zeta} using the trapezoid rule in order to obtain:

\begin{equation}
	\boxed{\zeta > \qty{1.255e43}{\per\solarmass\per\second}}
\end{equation}

\par
\subsection{Results}\label{sec:sim_results}
\begin{figure}
	\begin{center}
		\includegraphics[width=0.8\linewidth]{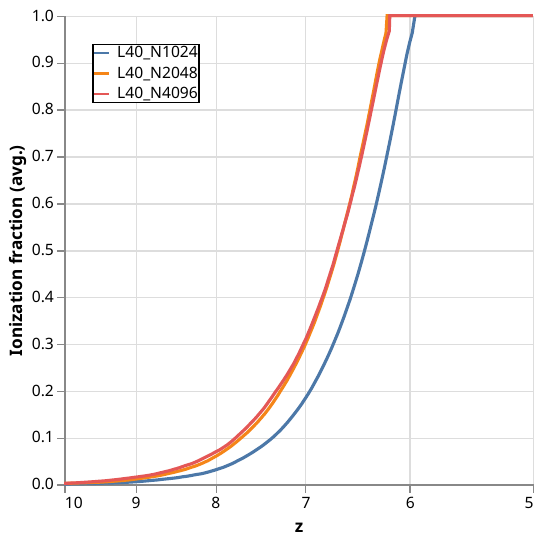}
	\end{center}
	\caption{Per-cell average ionized fraction vs.~redshift for results at $1024^3$, $2048^3$, and $4096^3$.  See Sec.~\ref{sec:sim_results}.}\label{fig:xh}
\end{figure}


\begin{table}[h]
  \centering
	\caption{Beginning, middle (avg. ionization fraction $> 50\%$) and end of reionization for our large-scale coupled runs (Sec~\ref{sec:large_scale_simulation}).}\label{table:reion_stats}
  \begin{tabular}{llll} 
    \toprule
    \multicolumn{1}{c}{Resolution} & \multicolumn{1}{c}{Start $z$} & \multicolumn{1}{c}{Mid $z$} & \multicolumn{1}{c}{End $z$} \\
    \midrule
    $1024^3$ & 11.798 & 6.38869 & 5.95168 \\
    $2048^3$ & 12.6478 & 6.66316 & 6.20206 \\
    $4096^3$ & 12.982 & 6.67314 & 6.17754 \\
    \bottomrule
  \end{tabular}
\end{table}

\begin{figure}
	\begin{center}
		\includegraphics[width=0.85\linewidth]{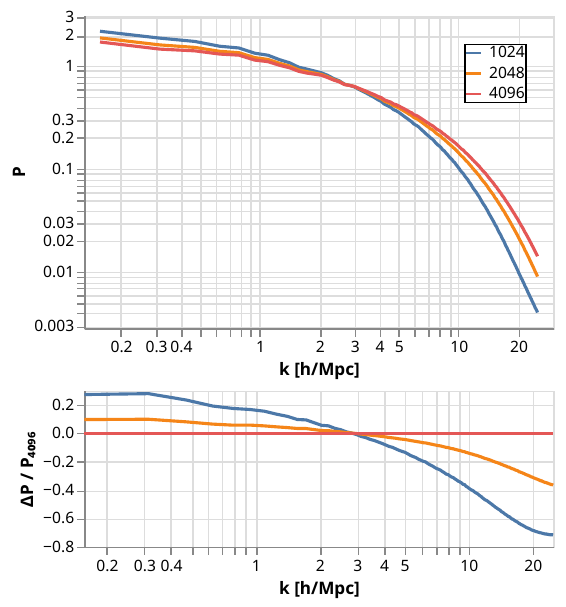}
	\end{center}
	\caption{Convergence of the 1D Lyman-alpha forest flux power spectrum at grid resolutions of $1024^3$, $2048^3$, and $4096^3$. Mean flux of all simulations has been normalized to 0.1439, which is the mean flux of $4096^3$ run. Top: Power $P(k)$ in units of \si{\per\h\mega\parsec}, where $k$ is the wavenumber. Bottom: Relative difference $\Delta P / P_{4096}$, where $P_{4096}$ is the value of $P(k)$ at $4096^3$, and $\Delta P = P_r - P_{4096}$, where $r \in \{1024, 2048, 4096\}$ is the resolution. See Sec.~\ref{sec:sim_results}.}\label{fig:power_spectrum}
\end{figure}

Plots of the temperature, density, and ionization fraction fields at $z=7$ at each resolution are shown in Fig.~\ref{table:slices7}. Qualitatively, ionization fronts at $2048^3$ and $4096^3$ appear closer to each other than to the $1024^3$ result, which is indicative of convergence and is corroborated by our quantitative measurements of average ionized fraction (Fig.~\ref{fig:xh}) and 1D Lyman-$\alpha$ forest flux (Fig.~\ref{fig:power_spectrum}). The temperature plots demonstrate our approximation for heating due to reionization (Sec.~\ref{sec:heating}).
\par
In Fig.~\ref{fig:xh}, we plot the evolution of the per-cell averaged ionized fraction as a function of redshift. Results are consistent with what we observe qualitatively in Fig.~\ref{table:slices7}: results converge as resolution increases. Redshifts of the start, middle, and end of reionization for each run are given in Table~\ref{table:reion_stats}.

\section{Conclusion} \label{sec:conc}

In this paper, we present a system for GPU ray tracing for numerical simulation of radiative transfer effects for modeling reionization events.
Our system is integrated with the \textsc{Nyx} $N$-body and hydrodynamics code and implemented using its native performance-portable \textsc{AMReX} infrastructure.
We demonstrate the robustness of our implementation by verifying against analytic solutions and numerical results from multiple, independent simulations.
\par
Our novel contributions include:
\begin{itemize}
	\item New formulation of adaptive ray tracing in terms of filters and transformations that can be used with \textsc{AMReX} particle abstractions, enabling GPU performance portability at scale.
	\item Coupling of radiation transport (via ray tracing) to \textsc{Nyx}'s hydrodynamics and $N$-body solvers.
	\item New source merging scheme to accelerate ray tracing computationally as reionization progresses.
	\item Novel solution for low-density neighbor cells when performing photon-conserving geometric overlap correction.
	\item Production-scale, Nyx-RT simulation on NERSC's Perlmutter supercomputer, using $4096^3$ cells and particles, and with ray tracing on a $1024^3$ coarse grid.
\end{itemize}
\par
While this paper demonstrates a functional system that can already be used for production-runs and studies of Lyman-$\alpha$ forest with improved modeling of reionization, it is still by no means an optimal system nor the final word on this topic.
We plan on replacing our current ad-hoc temperature jump (Sec.~\ref{sec:heating}) with a principled photoheating model. Additionally, the \cTwoRay{} technique that we adopt for integration over time and frequency only accounts for hydrogen. Potential solutions include extensions of the \cTwoRay{} technique that support helium reionization, such as the work of~\cite{Friedrich_2012}.
\par
Performance-wise, there is a space for further GPU optimizations. For instance, we do not explicitly consider whether memory access patterns are aligned and coalesced; improvements are likely necessary in order to fully utilize the memory bandwidth of the GPU.
In addition to optimizing on-node GPU performance, scalability could be significantly enhanced by implementing ray merging in the vein of~\cite{wise_enzomoray_2011} and \cite{Cain_2024}.
Furthermore, source grouping can be perhaps be improved further with octree structures featured in works on reverse ray tracing~\citep{Razoumov2002,Hasegawa2010,Okamoto2012,wunsch_tree-based_2018, grond_trevr_2019}.
Efficient GPU implementation of such structures can be nontrivial and is left to future work.
Finally, this paper focuses on ray tracing on a uniform grid, as that is the regime most relevant for our Lyman-$\alpha$ simulations. More extensive tests of ray tracing using the block-structured AMR paradigm of \textsc{AMReX} remains an important area of future work.
\par
In summary, we believe that the new method introduced in this work is an important step forward to accurately modeling the reionization processes in cosmological hydrodynamical simulations.
This method currently balances significant extra computing cost (compared with the standard approach used in IGM and Lyman-$\alpha$ studies) with realistic modeling and coupling of RT and hydrodynamics.
This represents a significant improvement over the previous inhomogeneous reionization model in Nyx \cite{onorbe_inhomogeneous_2019}.
As the raw computing power available to science applications increases, we plan to further develop this approach in the direction of higher physical fidelity using more accurate photo-heating model.

\section*{Acknowledgments}
Authors would like to thank Kenta Soga and Hidenobu Yajima (University of Tsukuba) for helpful discussions and preliminary cross-checks. K.S. would like to thank Joseph Hennawi and the ENIGMA group at UCSB for helpful discussions, guidance, and feedback.
\par
This research was in part supported by the Exascale Computing Project (17-SC-20-SC), a joint project of the U.S.
Department of Energy’s Office of Science and National Nuclear Security Administration, responsible for delivering a capable Exascale ecosystem, including software, applications, and hardware technology, to support the nation’s Exascale computing imperative.
\par
This work also supported by the Intel OneAPI CoE, the Intel Graphics
and Visualization Institutes of XeLLENCE, the DOE Ab-initio Visualization
for Innovative Science (AIVIS) grant 2428225, and the Data Science at Scale Summer School program at Los Alamos National Laboratory.
\par
K.S.~gratefully acknowledges support by the U.S.~Department of Energy, Office of Science, Office of Advanced Scientific Computing Research, Department of Energy Computational Science Graduate Fellowship under Award Number DE-SC0024386.
\par
This research used resources of the National Energy Research Scientific Computing Center, a DOE Office of Science User Facility supported by the Office of Science of the U.S.
Department of Energy under Contract No.~DEC02-05CH11231.
This research also used resources of the Oak Ridge Leadership Computing Facility at the Oak Ridge National Laboratory, which is supported by the Office of Science of the U.S.
Department of Energy under Contract No.~DE-AC05-00OR22725.



\normalsize
\bibliography{references}

\end{document}